\documentclass{emulateapj}
\usepackage{graphicx}
\usepackage{txfonts}
\newcommand{\Cu}[5]{\mbox{$#1\,^#2{\rm #3}^{{\rm #4}}_{\rm #5}$}}

\newcommand{\SH}{$S_{\!\!\rm H}$}

\begin{document}
\shorttitle{Statistical equilibrium of copper in the atmospheres
of metal-poor stars}\shortauthors{Shi et al.}

\title{\textbf{NLTE analysis of copper lines in different stellar populations}\altaffilmark{$\ast$}}
\altaffiltext{$\ast$}{Based on data obtained from ESO Science Archive and the Subaru telescope.}

\author{J. R. Shi\altaffilmark{1,2}, H. L. Yan\altaffilmark{1,2}, Z. M. Zhou\altaffilmark{1,2}, G. Zhao\altaffilmark{1,2}}
\affil{\altaffilmark{1}Key Laboratory of Optical Astronomy, National Astronomical
Observatories, Chinese Academy of Sciences, Beijing 100012, P. R.
China\\
\altaffilmark{2}School of Astronomy and Space Science, University of Chinese Academy of Sciences, Beijing 100049, China} \email{sjr@bao.ac.cn}

\date{Accepted on June 5, 2018 by ApJ}

\begin{abstract}
The copper abundances of 29 metal-poor stars are determined based on the high resolution, high signal-to-noise ratio spectra from the UVES spectragraph at the ESO VLT telescope. Our sample consists of the stars of the Galactic halo, thick- and thin-disk with [Fe/H] ranging from $\sim-$3.2 to $\sim0.0$\,dex. The non-local thermodynamic equilibrium (NLTE) effects of \ion{Cu}{1} lines are investigated, and line formation calculations are presented for an atomic model of copper including 97 terms and 1089 line transitions. We adopted the recently calculated photo-ionization cross-sections of \ion{Cu}{1}, and investigated the hydrogen collision by comparing the theoretical and observed line profiles of our sample stars. The copper abundances are derived for both local thermodynamic equilibrium (LTE) and NLTE based on the spectrum synthesis methods. Our results show that the NLTE effects for \ion{Cu}{1} lines are important for metal-poor stars, in particular for very metal-poor stars, and these effects depend on the metallicity. For very metal-poor stars, the NLTE abundance correction reaches as large as $\sim +0.5$ dex compared to standard LTE calculations. Our results indicate that [Cu/Fe] is under-abundant for metal-poor stars ($\sim -0.5$ dex) when the NLTE effects are included.

\end{abstract}


\section{Introduction}

The history of the chemical composition of our Galaxy is dominated by the nucleosynthesis occurring in many generations of stars. Metal-poor stars represent one of the main diagnostic tools to probe the early phases of the chemical evolution of our Galaxy. The variation in the elemental abundance ratios derived at different metallicites can be compared with the yields from supernovae (SNe) of different masses to check which ones have contributed to the Galactic chemical enrichment and when. Here, the preliminary result on the iron-group element copper is presented, the main goal is to better constrain its nucleosynthetic origin.

Several sources of Cu have been discussed: (i) in massive stars the weak s-process operates during core-helium and carbon-shell hydrostatic burning phases, as well as in explosive complete Ne burning phase \citep{WW95,PGH10,LC03}; (ii) in low mass stars the main s-process occurs during the asymptotic giant branch (AGB) \citep{AKW99}, and (iii) in long-lived type Ia supernovae explosive nucleosynthesis happens \citep{IBN99,THR04,FKS14}.

From the observational point of view, Cu abundances have been discussed by \citet{GS88} and \citet{SC88}, a secondary-like process for Cu (one requiring iron seeds from previous stellar generations, giving rise to an enrichment proportional to the iron content) was found by them \citep[also see][]{SGC91}. Further abundance determinations for this
elements in halo and disk stars were provided by \citet{PBS00}, and their results can be represented by a flat distribution [Cu/Fe] $=-0.75\pm 0.2$\,dex for low metallicity stars up to [Fe/H] $<-1.8$, followed by a linear increase with a slope close to 1 in the metallicity range $-1.5 <$ [Fe/H] $< -1$ \citep{RL08,IAC03}. While for the Galactic disk stars, there is a bending of the [Cu/Fe] distribution, and a distinct and separated trends are seen between thick- and thin-disk stars \citep[see also ][]{MKS02,RTL03,YSZ15,ZG16,MLR17,DTA17}. For very low-metallicity stars, Cu abundances have been derived by \citet{BIR04} and \citet{LBJ08} under the assumption of LTE from the near-UV lines of Cu I at 3247.53 and 3273.95\,\AA\ taking the effects of hyperfine structure(HFS) and isotopic splitting into account, and they found a plateau ($<$[Cu/Fe]$>\simeq -0.98$\,dex) at [Fe/H] $< -2.5$. The copper abundances for the bulge stars have been derived by \citet{JRK14}, and it is found that the trend of [Cu/Fe] ratios with [Fe/H] in the bulge is very different from the Galactic thin- and thick-disk stars. In the bulge the Cu abundance increases monotonically from [Cu/Fe] $= -0.84$\,dex in the most metal-poor star to [Cu/Fe] $\sim +$0.40\,dex in the most metal-rich stars. 
The copper abundance in globular clusters have been discussed in detail by \citet{SSI03}, they derived the copper abundances of 117 giants in 10 globular clusters (M\,3, M\,4, M\,5, M\,10, M\,13, M\,17, NGC\,7006, NGC\,288 and NGC\,362), and noted that the copper abundances in globular clusters appear to follow the trends found in the field. Similar conclusion have been reported for M\,28 \citep{VMM17}, M\,80 \citep{CBG15}, NGC\,4833 \citep{CBG14}, and NGC5897 \citep{KM14}. While it is reported that the copper abundances in the massive Galactic globular cluster $\omega$ Centauri fall below the corresponding mean ratio in the field stars by roughly 0.5\,dex \citep{CSS02,PPH02,SSC00}, maybe also in the globular cluster Ruprecht\,106 \citep{VGC13}.

\citet{SVT03} measured Cu abundances in a total of 12 stars across four dwarf spheroidal galaxies, e.g. Sculptor, Fornax, Carina, and Leo I, all but one of the stars were below [Fe/H]$<-1.0$. Their results indicate that the [Cu/Fe] ratio is constant ([Cu/Fe]$\sim -0.7$) for [Fe/H]$<-1.0$, while the metal-rich Fornax star has a Cu enhancement. \citet{MS05} derived Cu abundances for 14 red giants in Sagittarius dwarf spheroidal galaxy (Sgr), and noted that compared to Milky Way stars the [Cu/Fe] ratio of Sgr stars is deficiency by $\sim$0.5\,dex \citep[also see][]{CBG10,MWM13}. On the other hand, \citet{SBB07} found that the [Cu/Fe] deficiencies increase with increasing [Fe/H], such that the [Cu/Fe] ratio of Sgr stars is near -1\,dex around solar iron abundance.  \citet{JIS06} derived Cu abundances of 10 red giant stars in four old globular clusters in the Large Magellanic Cloud (LMC), and found that their behavior follows the stars in $\omega$ Cen with similar [Fe/H]. \citet{PHS08} found that in the inner disk LMC stars, the copper distribution is flat with a mean value of [Cu/Fe] $= -0.68$ \,dex, while, around the higher metallicity range the LMC stars present a clear under-abundance with respect to the Galactic disk. \citet{CBC12} determined the detail abundances of 22 elements including copper for eight clusters in LMC, they also noted the depleted [Cu/Fe] at high metallicity. Very recently, \citet{SMW17} investigated two stars in the LMC globular cluster NGC\,1718, and ascertained these two stars strongly deficient in copper.

It is noted by \citet{BCH10} that the \ion{Cu}{1} resonance lines are not reliable abundance indicators, and departures from LTE should be taken into account to properly describe these lines for both dwarfs and giants. Recently, \citet{RSL14} found a large difference of copper abundance ($\sim 0.56$\,dex) derived from \ion{Cu}{1} resonance and \ion{Cu}{2} lines \citep[also see][]{RL12}, thus, they suggested that \ion{Cu}{1} lines may not be formed in LTE, and clearly more work is needed to better understand the formation of \ion{Cu}{1} lines in cool stars.

The present work is based on a sample of metal-poor stars and aims at exploring their [Cu/Fe] abundance ratios applying full spectrum synthesis based on level populations calculated from the statistical equilibrium equations. In Sect. 2 we provided the observational techniques, while the atmospheric models, stellar parameters and atomic data are described in Sect. 3. The copper atomic model and the NLTE effects are discussed in Sect. 4. The results and the comparisons with other works are illustrated in Sec. 5. The discussion is given in Sect. 6, and the conclusions are presented in the last section.

\section{Observations}

Our aim is to derive the copper abundances for a sample of metal-poor stars using high resolution and high signal-to-noise ratio spectra with the two \ion{Cu}{1} near-UV resonance lines included. The spectra of 26 metal-poor stars were observed with the UVES \'{e}chelle spectrograph mounted at the ESO VLT during two observation runs: April 8-12 (2000) and April 10-12 (2001, programme IDS 65.L-0507 and 67.D-0439). The wavelength ranges of the spectra are from 3050 to 3850 \AA\ with a resolution power (R) of 48\,000 for the blue arm, while from 4800 to 6800 \AA\ with R $\sim$ 55\,000 for the red arm. We use also high-quality observed spectra from the ESO UVESPOP survey \citep{BJL05} for Procyon, HD\,84937 and HD\,122563. Finally, for G\,64-12 we use the spectrum from the High Dispersion Spectrograph at the Nasmyth focus of the Subaru 8.2 m telescope \citep{NAK02}.

The spectra were reduced with the standard ESO MIDAS package including location of \'{e}chelle orders, wavelength calibration, background subtraction, flat-field correction, and order
extraction.

\section{Method of calculation}

\subsection{Model atmospheres}

In this analysis the line-blanketed 1D LTE MAFAGS opacity sampling model atmospheres are adopted \citep{G04,GKT09}, and the convection according to \citet{CM92} is used. For stars with [Fe/H] $< -0.6$ the enhanced $\alpha$-element (O, Mg, Si and Ca) by 0.4\,dex  were used for individual models. As usual, the mixing length parameter l/Hp = 0.5 is adopted.

\subsection{Stellar parameters}

We employed the stellar parameters determined by \cite{TSZ09} for most of our program stars except G\,20-24 and G\,183-11, the surface gravities of these two stars have been revised using the parallaxes from \textsl{Gaia} DR1 \citep{Gaia16}. For HD\,61421, HD\,84937 and HD\,122563 the parameters were taken from \cite{MGS11} while for G\,64-12 from \cite{SGM09}. We also adopted the parameters from \cite{MJP17} for HD\,122563. In these works the wings of the Balmer lines have been used to derive the effective temperatures, and the HIPPARCOS parallaxes are adopted to determine the surface gravities. The iron abundances have been determined with the \ion{Fe}{2} lines, and the microturbulence velocities are estimated by requesting that the iron abundance derived from \ion{Fe}{2} lines should not depend on equivalent widths. The uncertainties for the temperature, surface gravity, metal abundance and microturbulence velocity are generally considered to be $\pm$80\,K, 0.1\,dex, 0.1\,dex and 0.2\,km s$^{-1}$ respectively.

\subsection {Atomic line data}

The relevant line data with their final solar fit \citep{SGZ14} $gf$ values are presented in Table \ref{table1}. The collisional broadening through van der Waals interaction with hydrogen atoms are calculated according to \cite{AO91,AO95} tables, and the HFS was included in our analysis with the data taken from \citet{B76}. Following \citet{AGS09} the isotopic ratio of $^{63}$Cu:$^{65}$Cu is adopted as 69\%:31\%.

\begin{table}
\caption[1]{Atomic data of copper lines.}
\centering{\begin{tabular}{rr@{ $-$ }lrr}
\hline\hline\noalign{\smallskip}
$\lambda$~ [\AA] & \multicolumn{2}{c}{Transition} & $\log gf$ & $\log C_6$\\
\hline\noalign{\smallskip}
 3247.54 &  \Cu{4s}{2}{S}{}{1/2}   & \Cu{4p}{2}{^{}P}{}{3/2} & -0.21 & -31.73  \\
 3273.96 &  \Cu{4s}{2}{P}{}{1/2}   & \Cu{4p}{2}{P}{}{1/2} & -0.50 & -31.74  \\
 5105.54 &  \Cu{4s^2}{2}{D}{}{5/2} & \Cu{4p}{2}{P}{}{3/2} & -1.64 & -31.67  \\
 5218.20 &  \Cu{4p}{2}{P}{}{3/2}   & \Cu{4d}{2}{D}{}{5/2} & +0.28 & -30.57  \\
 5220.07 &  \Cu{4p}{2}{P}{}{3/2}   & \Cu{4d}{2}{D}{}{3/2} & -0.63 & -30.57  \\
 5700.24 &  \Cu{4s^2}{2}{D}{}{3/2} & \Cu{4p}{2}{P}{}{3/2} & -2.60 & -31.65  \\
 5782.13 &  \Cu{4s^2}{2}{D}{}{3/2} & \Cu{4p}{2}{P}{}{1/2} & -1.89 & -31.66  \\

\noalign{\smallskip} \hline
\end{tabular}}
\label{table1}
\end{table}

\section{NLTE calculations}

\subsection{Atomic model}

Our copper model atom contains all the important \ion {Cu}{1} levels (including 96 \ion {Cu}{1} terms) and the \ion{Cu}{2}
ground state, and is discussed in detail in \citet{SGZ14}. Recently, \cite{LCZ14} have calculated the oscillator strengths and photoionization cross-sections using the R-matrix method in the LS-coupling scheme, and we adopted their results in our atomic model. Following \cite{SGZ14} the hydrogen collision enhancement factor \SH\ = 0.1 is taken in our analysis.

A revised version of the DETAIL program \citep{BG85} have been adopted to solve the coupled radiative transfer and statistical equilibrium equations, which are based on the
accelerated lambda iteration following the method described by \citet{RH91,RH92}, and it was presented in detail by \citet{MGS11}. The obtained departure coefficients were then adopted with the SIU code \citep{R91} to calculate the synthetic line profiles.

\subsection{NLTE effects}

\begin{figure}
\resizebox{\hsize}{6.6cm}{\includegraphics[width=9.3cm]{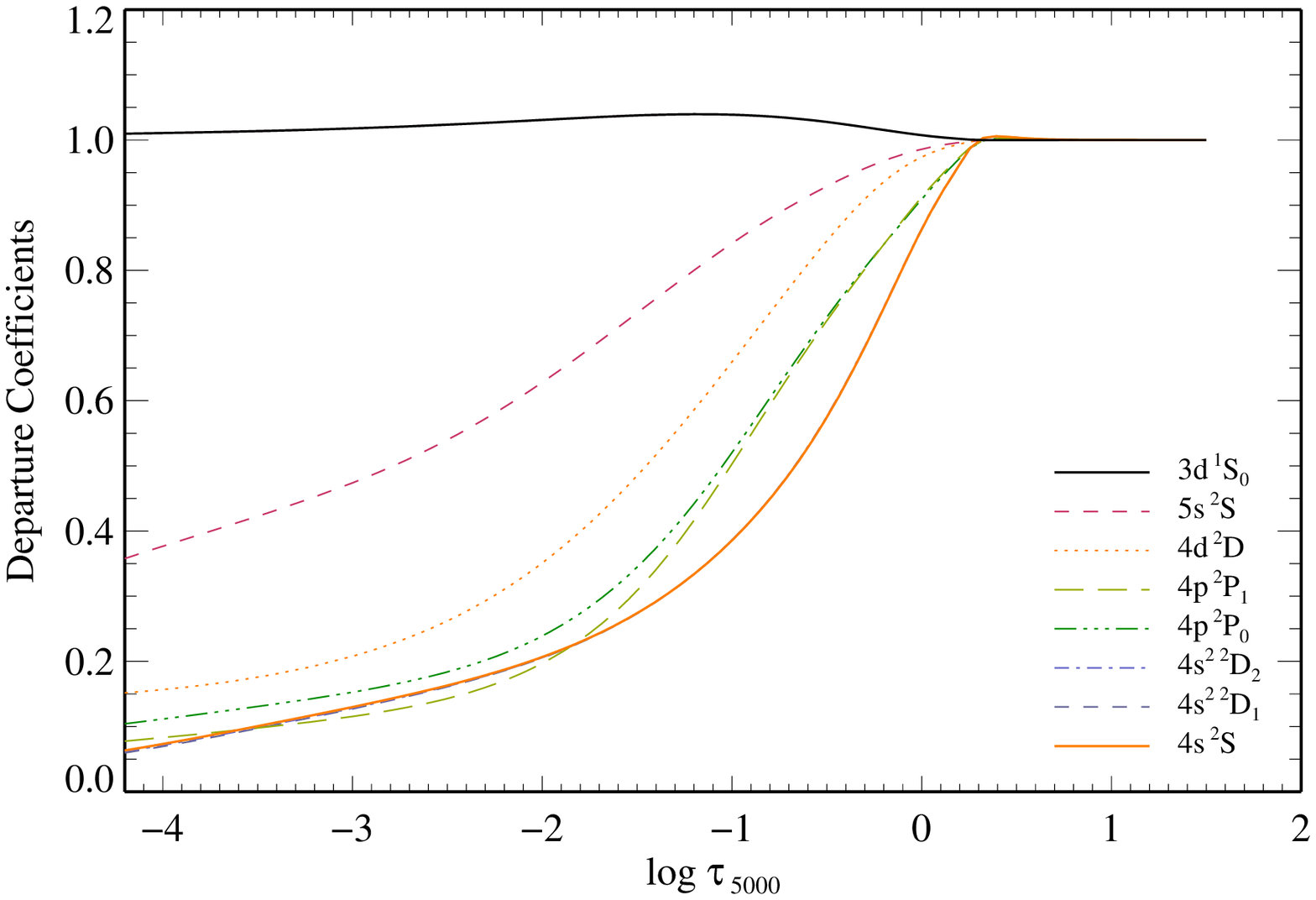}}
\resizebox{\hsize}{6.6cm}{\includegraphics[width=9.3cm]{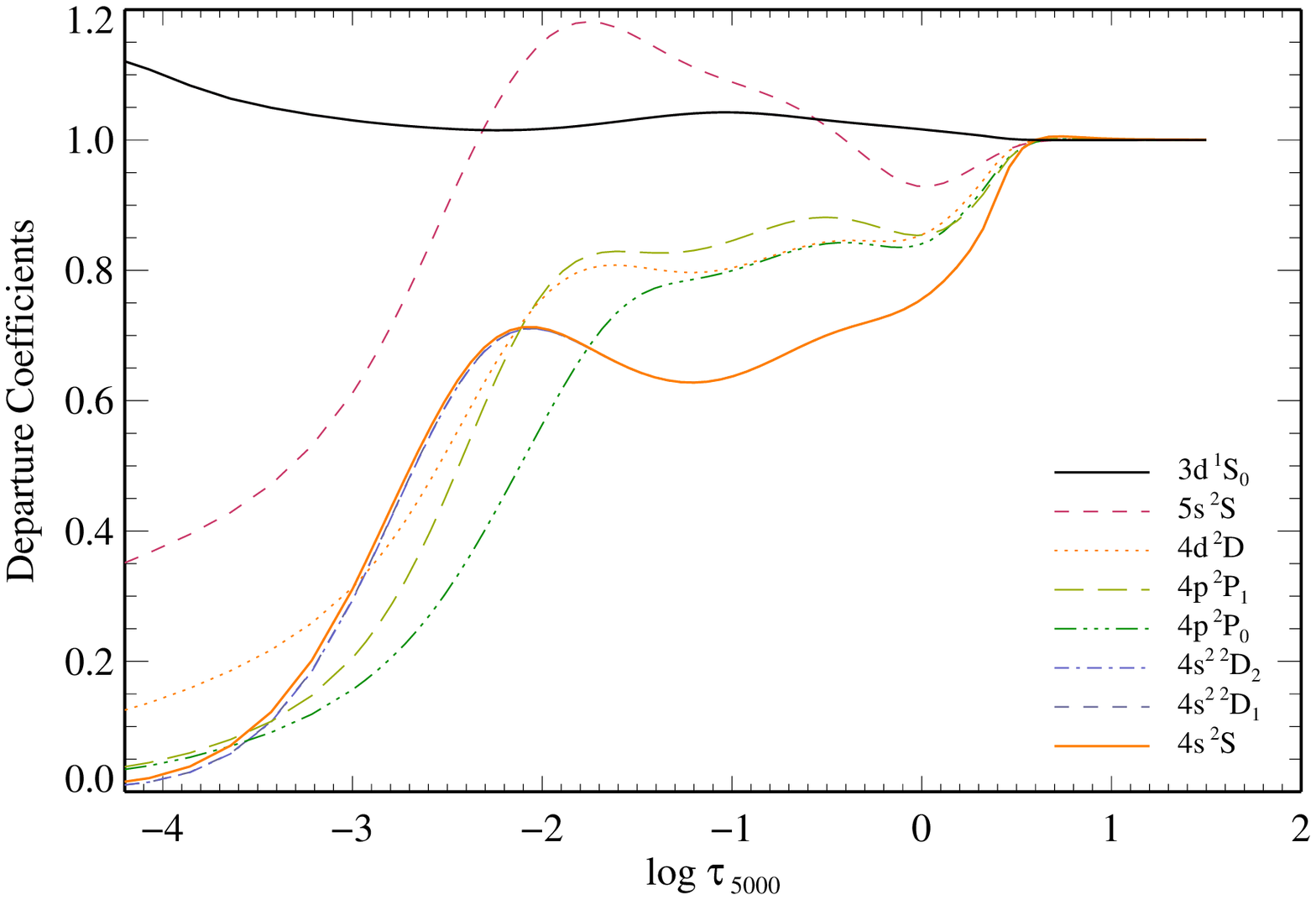}}
\caption[short title]{Departure coefficients ($b_i$) for investigated energy \ion{Cu}{1} levels and the \ion{Cu}{2} ground state as a function of continuum optical depth at 5000\,\AA\ for the model atmospheres of HD\,84937 (upper) and HD\,122563 (bottom). The neutral hydrogen collision factor is scaled by 0.1.}
\vspace*{-0.0cm} \label{fig1}
\end{figure}

The departure coefficients ($b_i = n_i^{NLTE}/n_i^{LTE}$) of the investigated levels as a function of the continuum optical depth at 5000\,\AA\ are shown in Fig.\,\ref{fig1} for the model atmospheres of HD\,84937 and HD\,122563, here $b_i$ is the ratio of the number density of NLTE to that of LTE. HD\,84937 is a well-known metal-poor star with high-temperature among our sample, while HD\,122563 is a typical cool metal-poor giant star. The departure coefficients for the investigated levels of \ion{Cu}{1} and the \ion{Cu}{2} ground state are presented in this figure, and it is clear that the number densities of the \ion{Cu}{1} levels begin to underpopulate outside layers with log $\tau_{5000} \sim 0.5$ due to over-ionization for both stars. An obvious overpopulate of the level ${5s}$ $^{2}{S}$ for HD\,122563, which is due to over-recombination, can be seen in the range of $-0.5 >$ log $\tau_{5000} > -2.0$.

\begin{figure}
\resizebox{\hsize}{6.6cm}{\includegraphics[width=9.3cm]{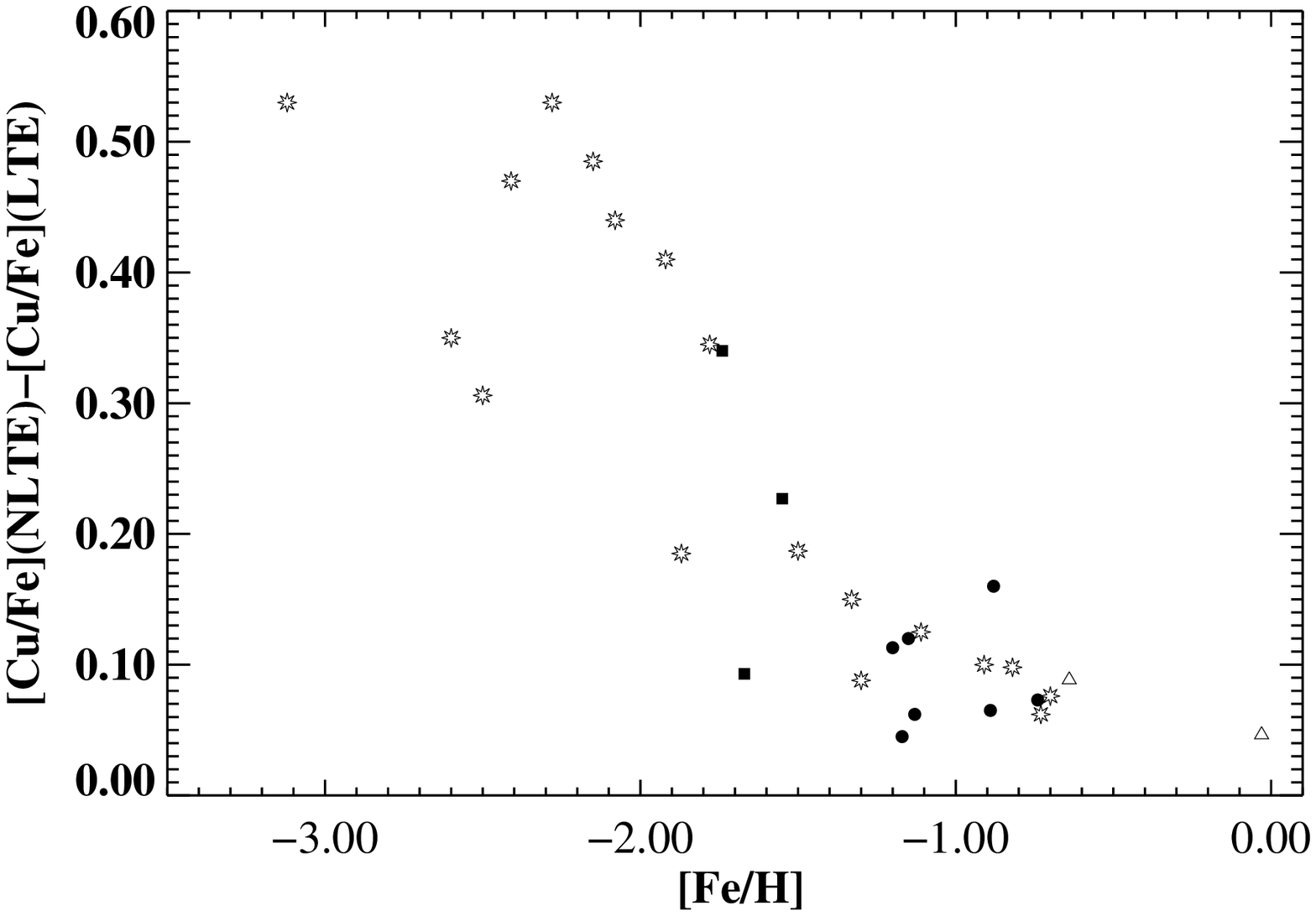}}
\resizebox{\hsize}{6.6cm}{\includegraphics[width=9.3cm]{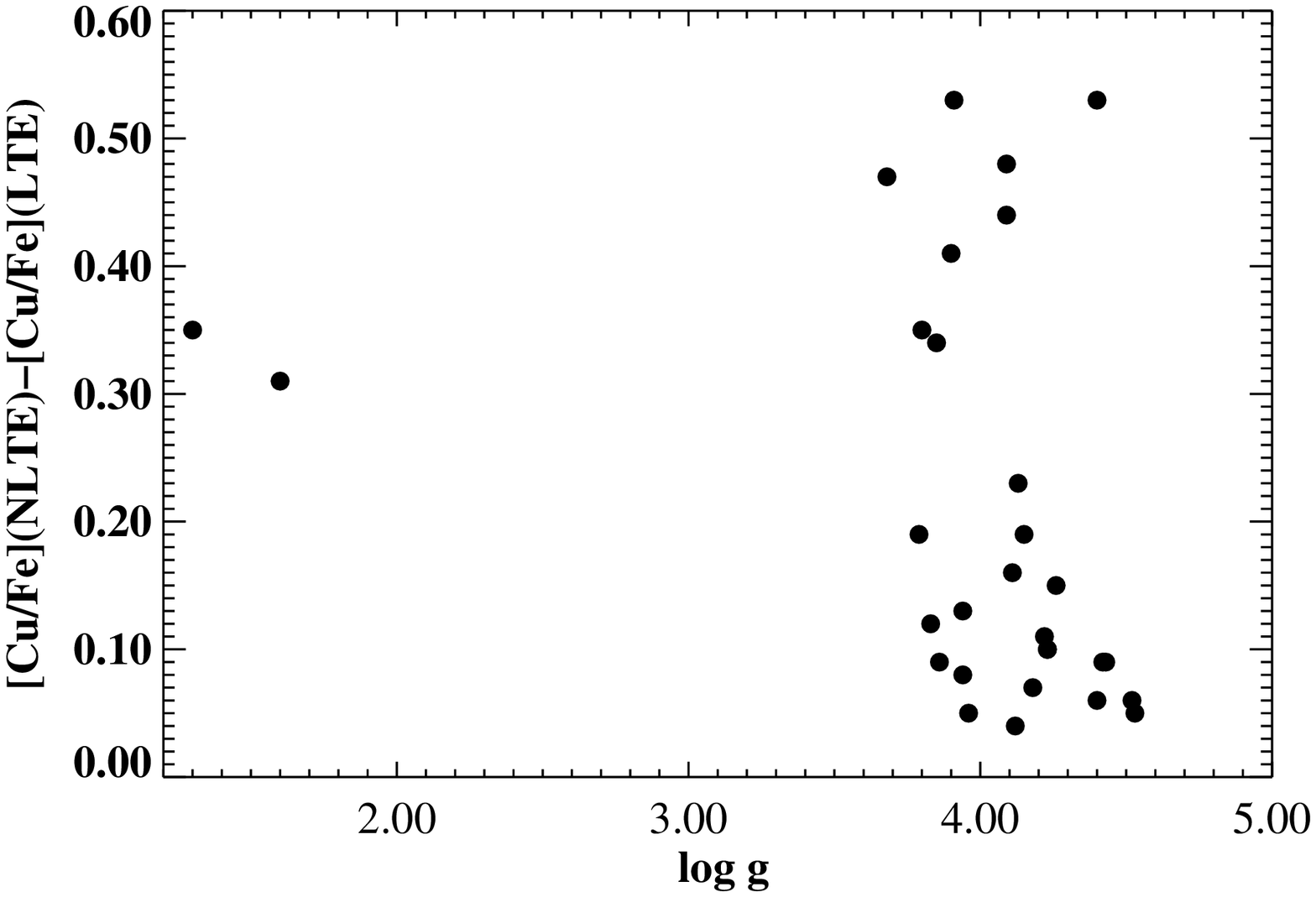}}
\resizebox{\hsize}{6.6cm}{\includegraphics[width=9.3cm]{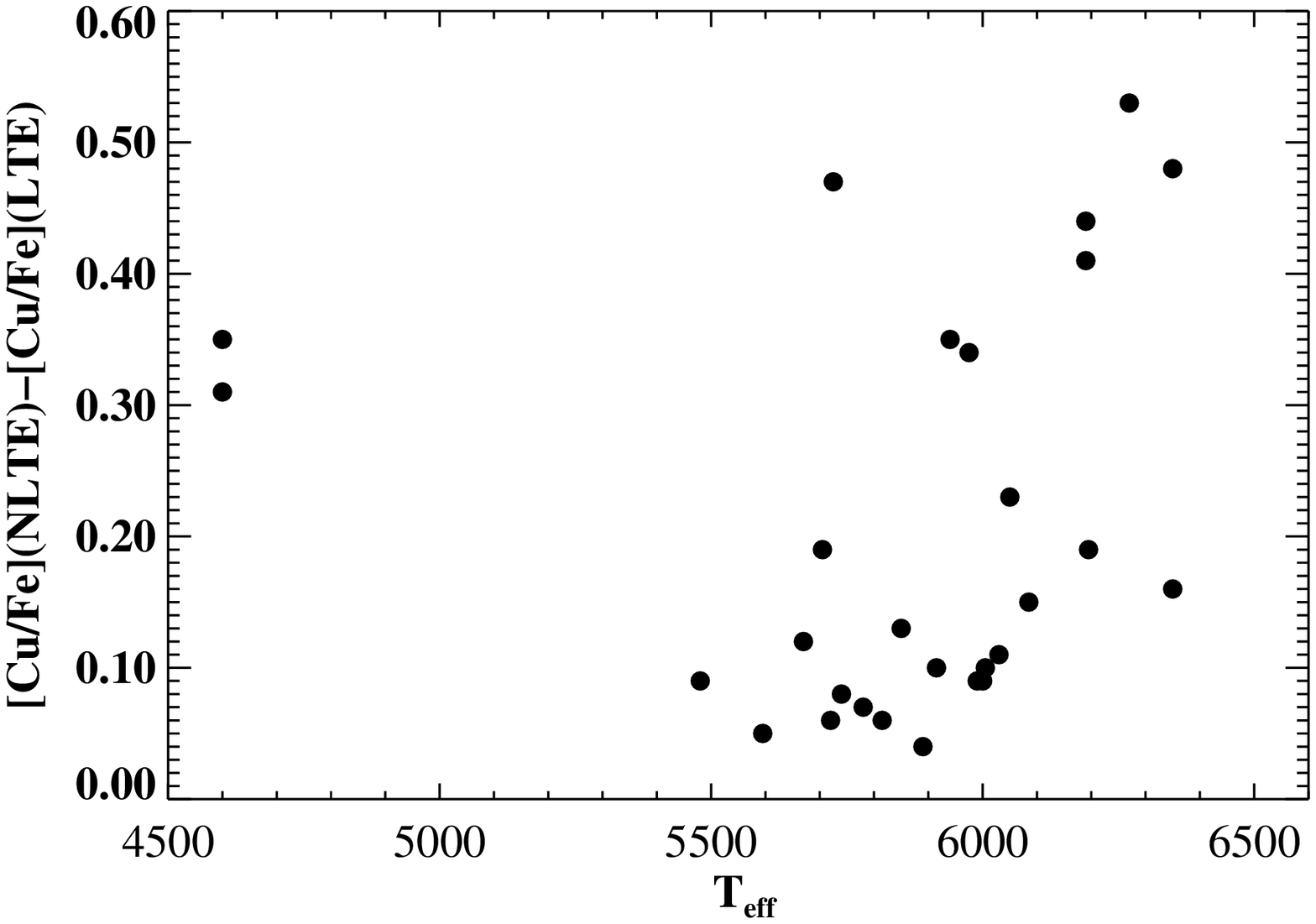}}
\caption[short title]{Difference of [Cu/Fe] abundance ratios
derived from LTE and NLTE cases as a function of metal
abundance (a), temperature (b), and surface gravity (c).}
\vspace*{-0.0cm} \label{fig2}
\end{figure}

It is noted that the NLTE effects of \ion{Cu}{1} lines are evident in our abundance determination, and as expected, there is a inclination that the effects tend to be large for more metal-poor stars, very similar phenomenon is discovered for sodium \citep{SGZ04}. Compared to their NLTE counterparts the substantially lower LTE results can be found clearly, and the differences can be larger than 0.50\,dex for extreme metal-poor stars. We display the copper abundance differences between LTE and NLTE calculations as functions of temperature, metal abundance and surface gravity for our sample stars in Fig. \ref{fig2}, respectively. The average NLTE effects are $+0.27$, $+0.09$ and $+0.07$\,dex for the halo, thick and thin disk stars.

Our results indicate that departures from LTE of the copper level populations seem to be larger for more metal-poor stars, and there is a clear tendency that the NLTE effects increase with decreasing metallicity, which can explain the large Cu abundance difference derived from \ion{Cu}{1} and \ion{Cu}{2} lines for stars of [Fe/H] $\sim -2.3$, e.g., the Cu abundances derived from \ion{Cu}{1} resonance lines are about 0.56\,dex lower than that from the \ion{Cu}{2} lines found by \citet{RSL14}.

\section{Results}

\subsection{Stellar copper abundances}

\begin{figure}
\resizebox{\hsize}{6.6cm}{\includegraphics[width=9.3cm]{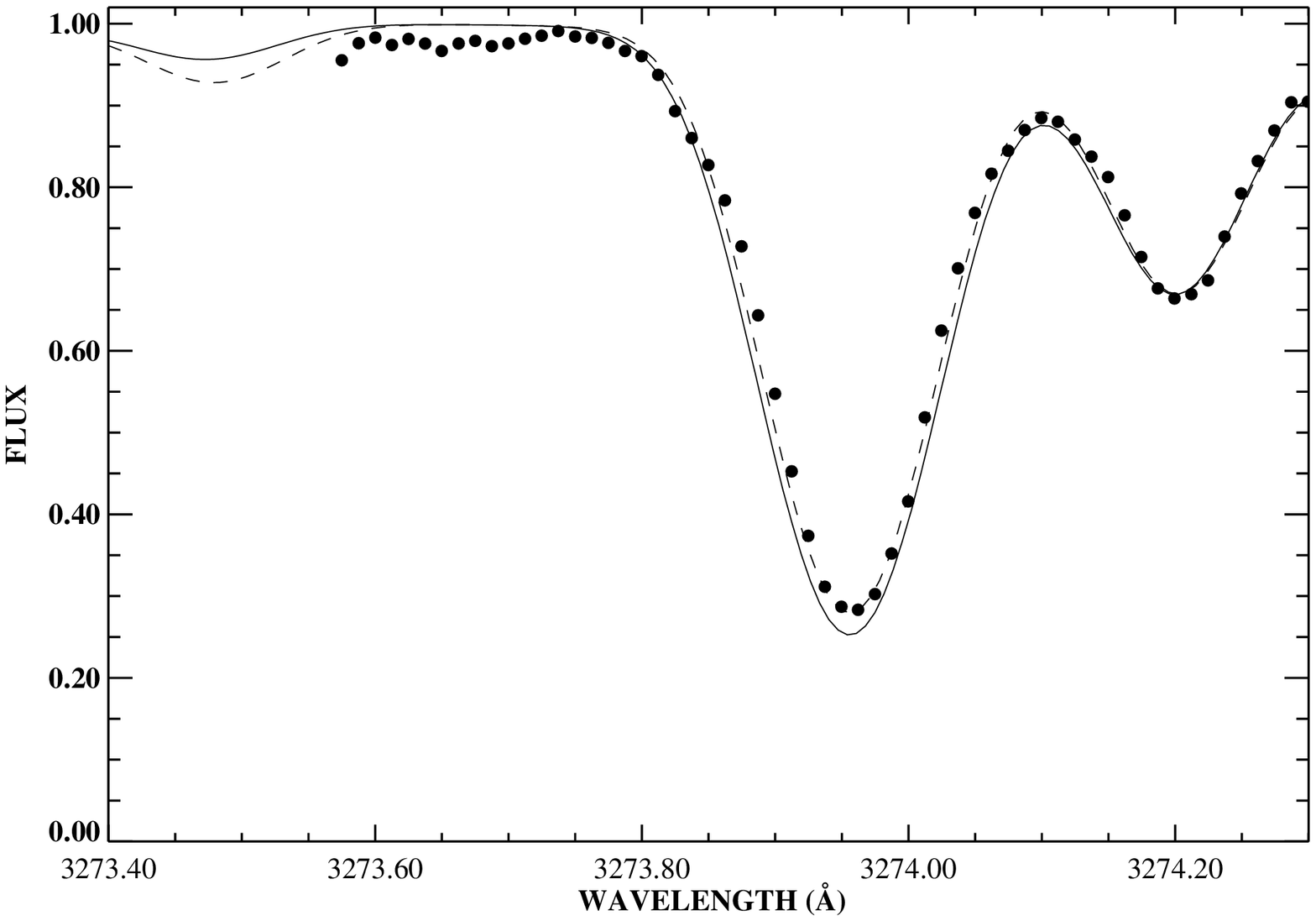}}
\resizebox{\hsize}{6.6cm}{\includegraphics[width=9.3cm]{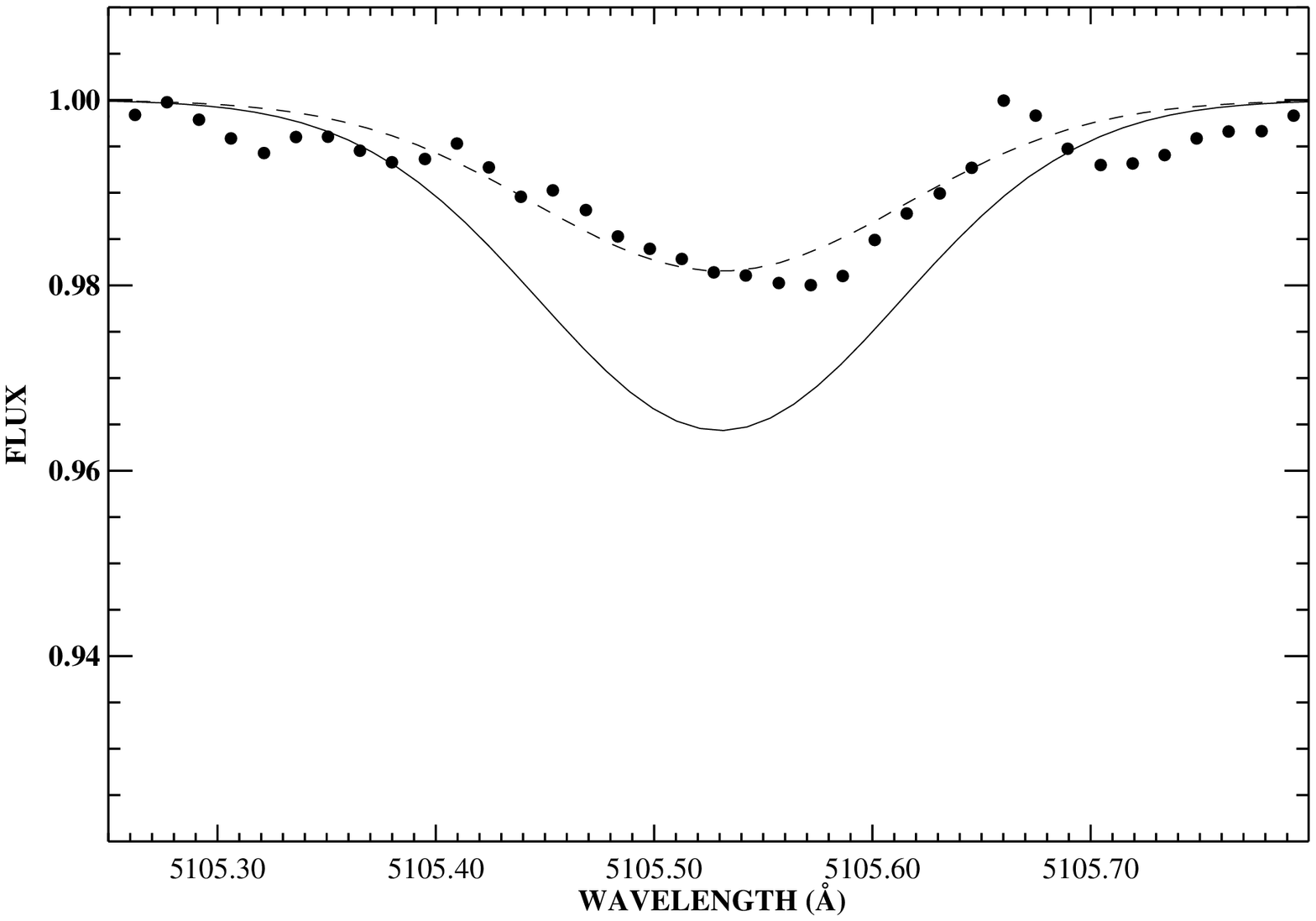}}
\caption[short title]{the observed \ion{Cu}{1} lines of HD\,122563 (filled circles), compared to the NLTE (dotted lines) and LTE (solid lines) synthetic profiles.}
\vspace*{-0.0cm} \label{fig3}
\end{figure}

\begin{figure}
\resizebox{\hsize}{6.6cm}{\includegraphics[width=9.3cm]{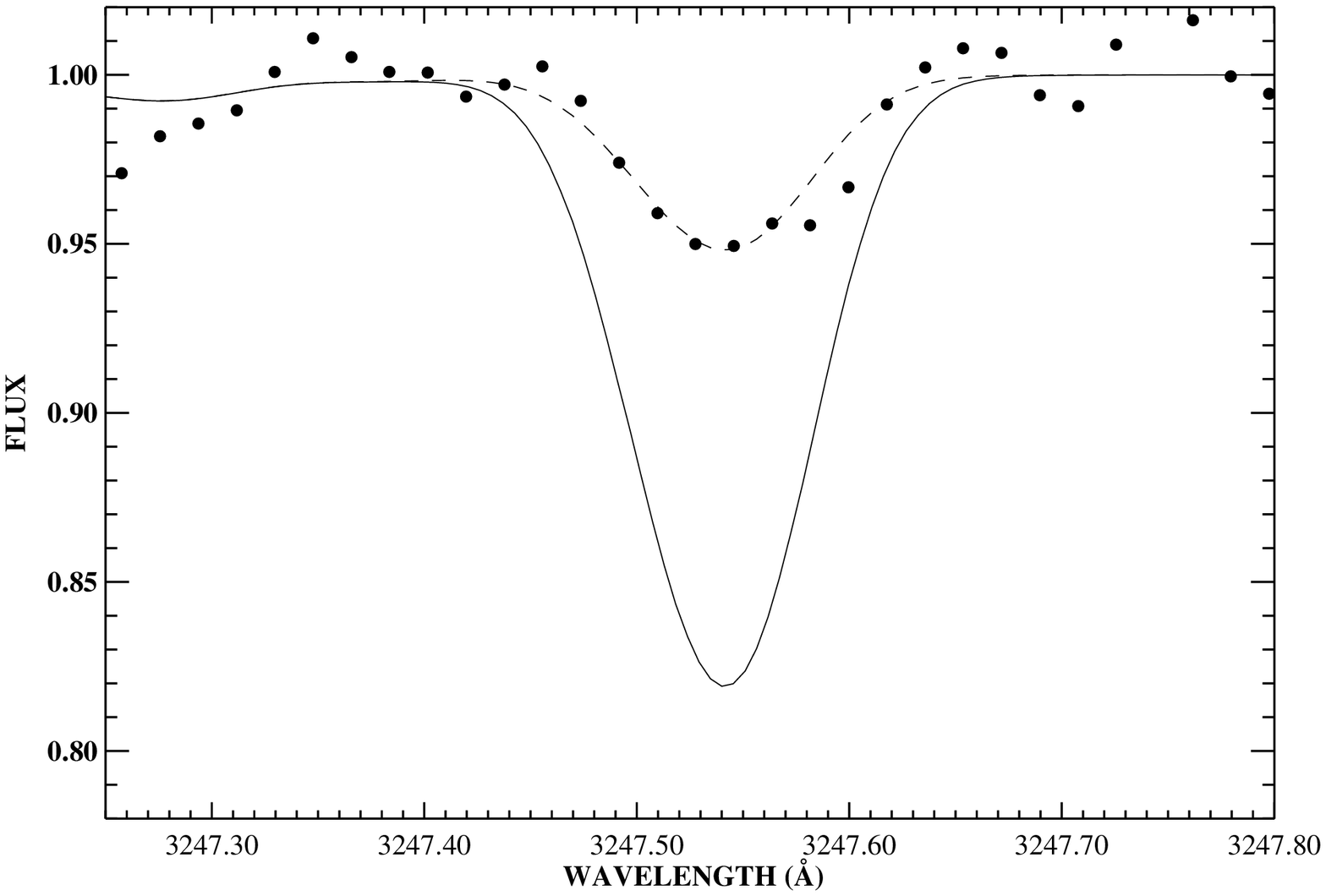}}
\caption[short title]{Similar to Fig. \ref{fig3} but for G\,64-12.}
\vspace*{-0.0cm} \label{fig4}
\end{figure}

The copper abundances of our program stars are derived with the spectral synthesis method, and the synthetic spectra are convolved with a Gauss broadening profiles in order to fit the observed spectral lines. Figs.\ref{fig3} and \ref{fig4} show the fitting line profiles for HD\,122563 and G\,64-12, respectively. It is found that the abundance differences are small for the NLTE results with a line to line abundance scatter between 0.01 and 0.11\,dex, while it is larger for the LTE result. In Table \ref{table2} we present the final LTE and NLTE copper abundance results.

As noted by \cite{BIR04} and \cite{CDS04}, the abundances derived from the spectra of metal-poor stars may be overestimated if the continuum scattering is included as an additional opacity source in the spectral synthesis code. The overestimation is especially high for lines with $\lambda <$ 4000\,\AA, where continuum scattering becomes important relative to continuous absorption. Based our spectrum synthesis modeling code SIU, which can treat continuum scattering properly, we evaluated the influence on the derived Cu abundances for the only giant, HD\,122563, and the determined Cu abundance lowers by about 0.3\,dex than that derived without the continuum scattering considered, while there is no impact on the derived Cu abundances for the dwarfs, e.g. HD\,84937 and G\,64-12. We found a much less correlation of Cu abundance with wavelength when the continuum scattering is included, similar behavior has been found by \cite{CDS04} and \cite{LBJ08}. 

\begin{figure}
\resizebox{\hsize}{6.6cm}{\includegraphics[width=9.3cm]{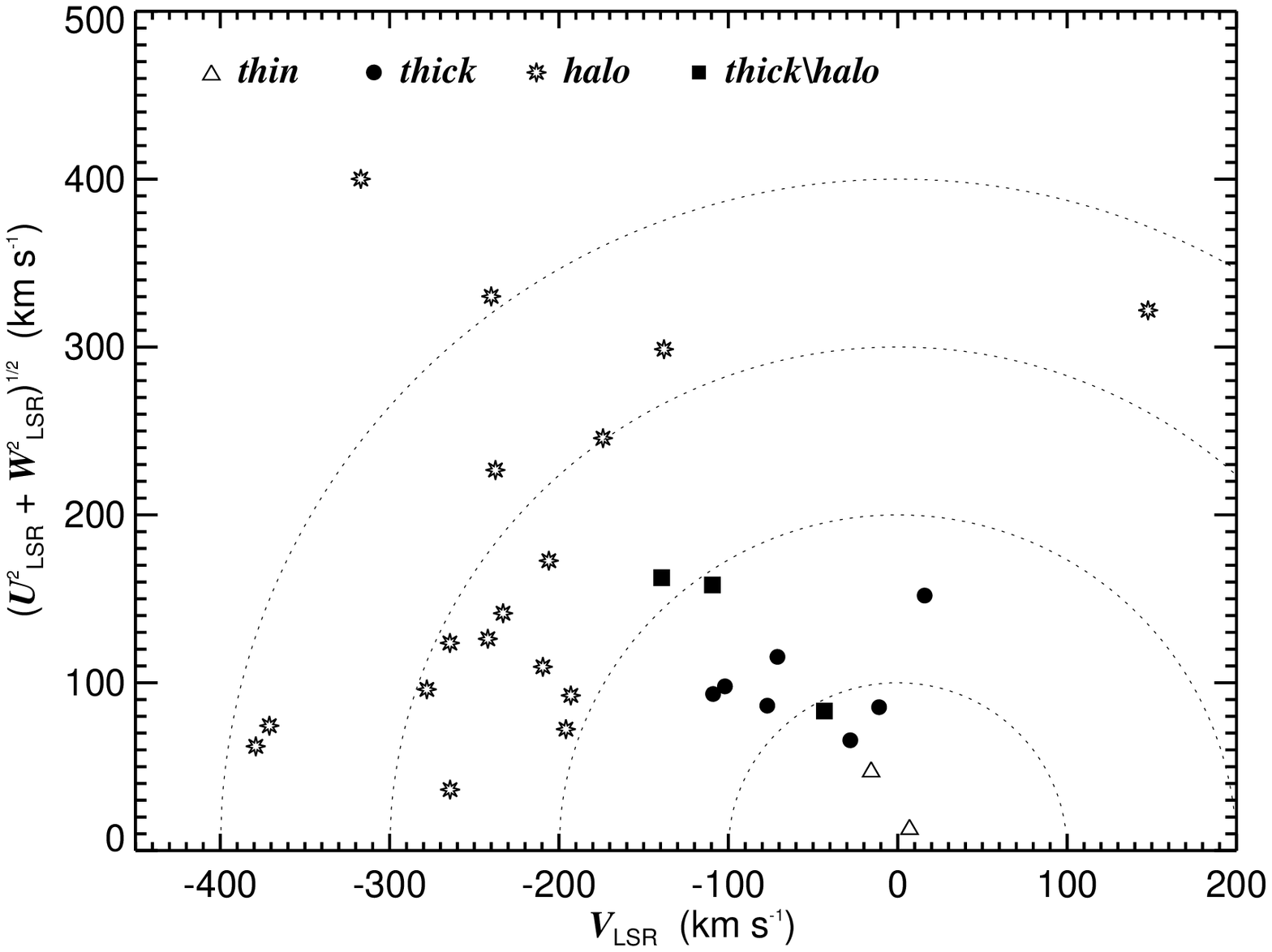}}
\caption[short title]{Toomre diagram of our program stars. Different symbols correspond
to different stellar populations, namely, the thin disk (equilateral triangle), thick disk (filled circle), thick/halo (filled square) and halo (Eight pointed star).} \label{fig5}
\end{figure}

\begin{figure}
\resizebox{\hsize}{6.6cm}{\includegraphics[width=9.3cm]{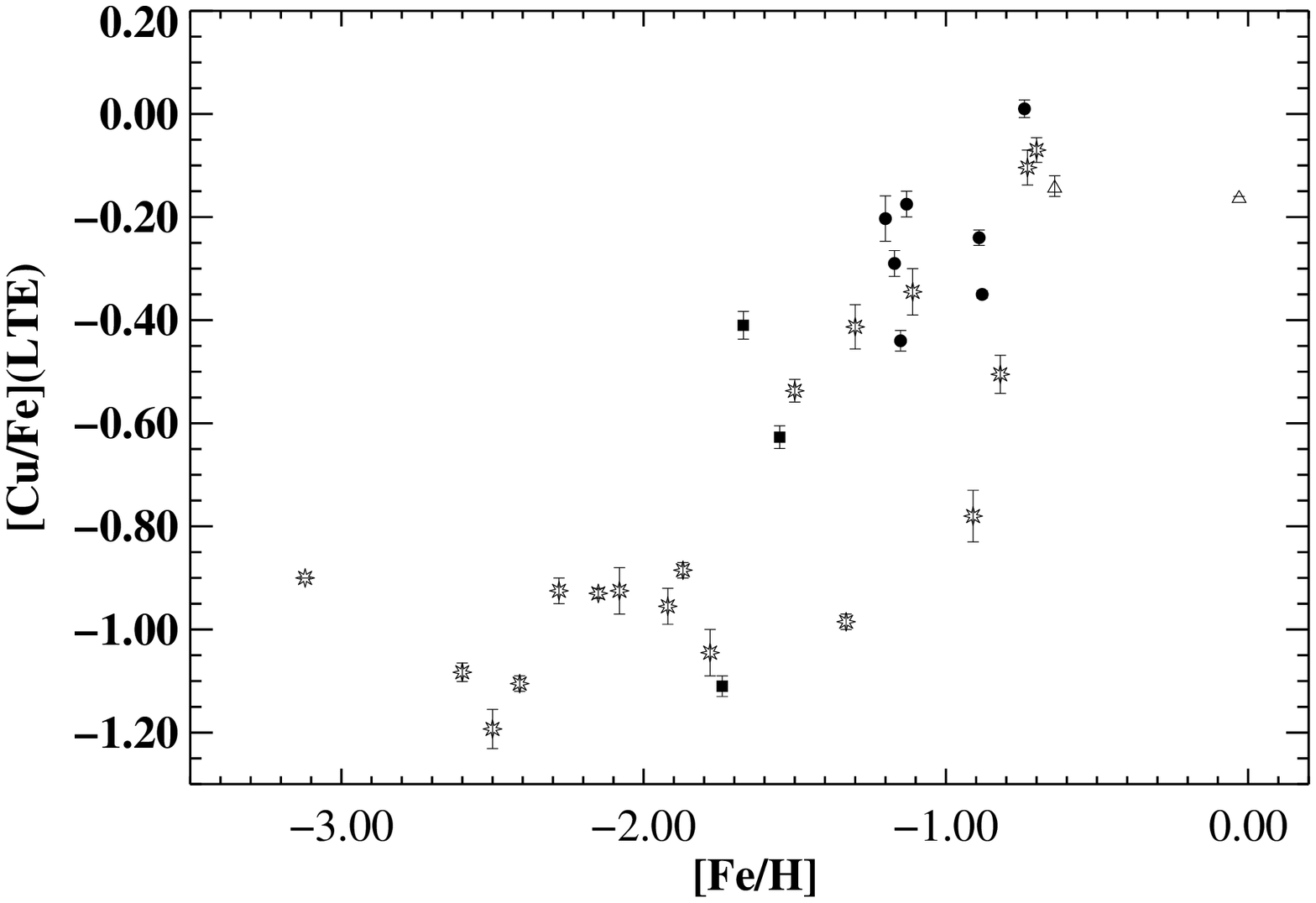}}
\caption[short title]{The [Cu/Fe] ratios under LTE situation as a function of [Fe/H] for selected stars. Symbols are same as Fig. \ref{fig5}.} \label{fig6}
\end{figure}

\begin{figure}
\resizebox{\hsize}{6.6cm}{\includegraphics[width=9.3cm]{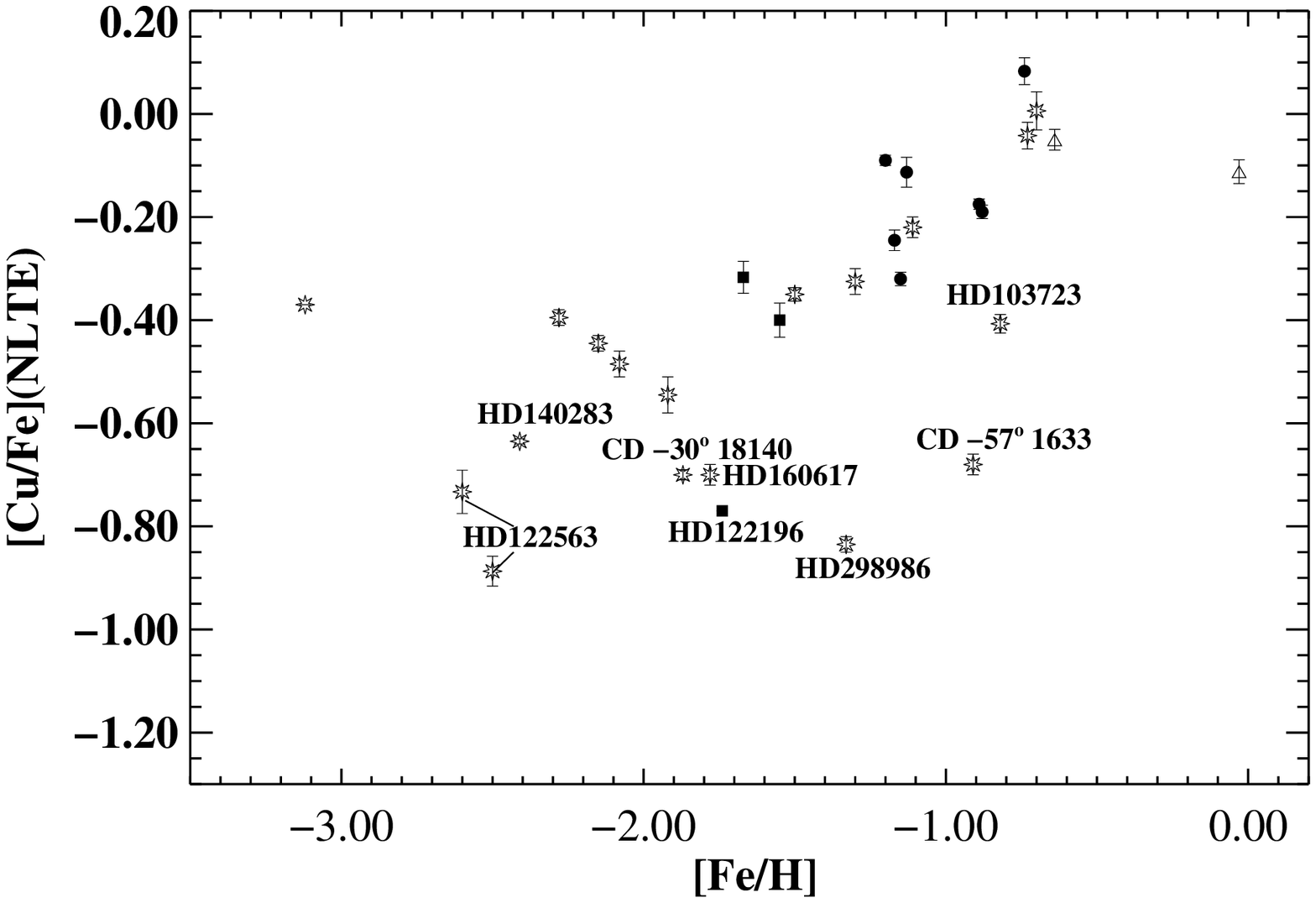}}
\caption[short title]{The [Cu/Fe] ratios under NLTE situation as a function of [Fe/H] for selected stars. Symbols are same as Fig. \ref{fig5}.} \label{fig7}
\end{figure}

\begin{table*}
\caption[2]{Atmospheric parameters and copper abundances of the selected stars. The LTE results of each star are shown at the first row, while the NLTE cases are indicated at the second row. The information of the stellar parameters are described in the text.}
\setlength{\tabcolsep}{0.1cm}
\centering
\begin{tabular}{lccclccccccrc}
\hline\hline\noalign{\smallskip}
   Name   & $T {\rm _{eff}}$  & $\log g$ & [Fe/H] &$\xi$   &   3247 & 3273  & 5105 & 5218 & 5220&  5700 & 5782 &[Cu/Fe]\\
\noalign{\smallskip}\hline\noalign{\smallskip}
CD\,$-$30$^\circ$ 18140& 6195 &4.15  & -1.87 &  1.5  & -0.87 & -0.90 &  --   &  --   &  --   &  --   &   --  & -0.89$\pm$0.015\\
                       &      &      &       &       & -0.69 & -0.71 &  --   &  --   &  --   &  --   &   --  & -0.70$\pm$0.010\\
CD\,$-$57$^\circ$ 1633 & 5915 &4.23  & -0.91 &  1.2  & -0.82 & -0.81 & -0.81 & -0.68 &  --   &  --   &   --  & -0.78$\pm$0.050\\
                       &      &      &       &       & -0.71 & -0.69 & -0.66 & -0.66 &  --   &  --   &   --  & -0.68$\pm$0.020\\
G\,13-009              & 6270 &3.91  & -2.28 &  1.5  & -0.95 & -0.90 &   --  &   --  &   --  &   --  &   --  & -0.93$\pm$0.025\\
                       &      &      &       &       & -0.41 & -0.38 &   --  &   --  &   --  &   --  &   --  & -0.40$\pm$0.015\\
G\,020-024             & 6190 &3.90  & -1.92 &  1.5  & -0.99 & -0.92 &   --  &   --  &   --  &   --  &   --  & -0.96$\pm$0.035\\
                       &      &      &       &       & -0.58 & -0.51 &   --  &   --  &   --  &   --  &   --  & -0.55$\pm$0.035\\
G\,64-12               & 6407 & 4.40 & -3.12 &  2.5  & -0.90 &   --  &   --  &   --  &   --  &   --  &   --  & -0.90$\pm$0.000\\
                       &      &      &       &       & -0.37 &   --  &   --  &   --  &   --  &   --  &   --  & -0.37$\pm$0.000\\
G\,183-011             & 6190 &4.09  & -2.08 &  1.5  & -0.97 & -0.88 &   --  &   --  &   --  &   --  &   --  & -0.93$\pm$0.045\\
                       &      &      &       &       & -0.51 & -0.46 &   --  &   --  &   --  &   --  &   --  & -0.49$\pm$0.025\\
HD\,61421              & 6510 & 3.96 & -0.03 &  1.8  &   --  &   --  & -0.14 & -0.17 & -0.15 &   --  & -0.18 & -0.16$\pm$0.015\\
                       &      &      &       &       &   --  &   --  & -0.07 & -0.13 & -0.11 &   --  & -0.14 & -0.11$\pm$0.023\\
HD\,76932              & 5890 &4.12  & -0.89 &  1.2  & -0.22 & -0.25 & -0.26 & -0.23 &  --   &  --   &   --  & -0.24$\pm$0.032\\
                       &      &      &       &       & -0.19 & -0.18 & -0.16 & -0.17 &  --   &  --   &   --  & -0.18$\pm$0.010\\
HD\,84937              & 6350 &4.09  & -2.15 &  1.7  & -0.92 & -0.94 &   --  &   --  &   --  &   --  &   --  & -0.93$\pm$0.010\\
                       &      &      &       &       & -0.43 & -0.46 &   --  &   --  &   --  &   --  &   --  & -0.45$\pm$0.015\\
HD\,97320              & 6030 &4.22  & -1.20 &  1.3  & -0.19 & -0.18 & -0.29 & -0.15 &  --   &  --   &   --  & -0.20$\pm$0.044\\
                       &      &      &       &       & -0.09 & -0.10 & -0.10 & -0.07 &  --   &  --   &   --  & -0.09$\pm$0.010\\
HD\,97916              & 6350 &4.11  & -0.88 &  1.5  & -0.34 & -0.36 & -0.35 &   --  &  --   &  --   &   --  & -0.35$\pm$0.007\\
                       &      &      &       &       & -0.17 & -0.20 & -0.20 &   --  &  --   &  --   &   --  & -0.19$\pm$0.013\\
HD\,103723             & 6005 &4.23  & -0.82 &  1.3  & -0.57 & -0.43 & -0.51 & -0.51 &  --   &  --   &   --  & -0.51$\pm$0.037\\
                       &      &      &       &       & -0.42 & -0.38 & -0.40 & -0.43 &  --   &  --   &   --  & -0.41$\pm$0.018\\
HD\,106038             & 5990 &4.43  & -1.30 &  1.2  & -0.47 & -0.38 & -0.44 & -0.36 &  --   &  --   &   --  & -0.41$\pm$0.043\\
                       &      &      &       &       & -0.37 & -0.33 & -0.32 & -0.28 &  --   &  --   &   --  & -0.33$\pm$0.025\\
HD\,111980             & 5850 &3.94  & -1.11 &  1.2  & -0.32 & -0.28 & -0.38 & -0.40 &  --   &  --   &   --  & -0.35$\pm$0.045\\
                       &      &      &       &       & -0.21 & -0.21 & -0.20 & -0.26 &  --   &  --   &   --  & -0.22$\pm$0.020\\
HD\,113679             & 5740 &3.94  & -0.70 &  1.2  & -0.06 & -0.08 & -0.06 & -0.12 & -0.03 &  --   &   --  & -0.07$\pm$0.024\\
                       &      &      &       &       & -0.01 &  0.02 &  0.07 & -0.07 &  0.02 &  --   &   --  &  0.01$\pm$0.037\\
HD\,121004             & 5720 &4.40  & -0.73 &  1.1  & -0.12 & -0.12 & -0.12 & -0.14 & -0.02 &  --   &   --  & -0.10$\pm$0.034\\
                       &      &      &       &       & -0.03 & -0.05 & -0.03 & -0.10 &  0.00 &  --   &   --  & -0.04$\pm$0.026\\
HD\,122196             & 5975 &3.85  & -1.74 &  1.5  & -1.13 & -1.09 &   --  &   --  &  --   &  --   &   --  & -1.11$\pm$0.020\\
                       &      &      &       &       & -0.77 & -0.77 &   --  &   --  &  --   &  --   &   --  & -0.77$\pm$0.000\\
HD\,122563             & 4600 & 1.60 & -2.50 &  1.9  & -1.25 & -1.18 & -1.15 &   --  &   --  &   --  &   --  & -1.19$\pm$0.038\\
                       &      &      &       &       & -0.88 & -0.93 & -0.85 &   --  &   --  &   --  &   --  & -0.89$\pm$0.029\\
                       & 4600 & 1.32 & -2.63 &  1.7  & -1.07 & -1.11 & -1.07 &   --  &   --  &   --  &   --  & -1.08$\pm$0.018\\
                       &      &      &       &       & -0.75 & -0.78 & -0.67 &   --  &   --  &   --  &   --  & -0.73$\pm$0.042\\
HD\,126681             & 5595 &4.53  & -1.17 &  0.7  & -0.32 & -0.24 & -0.31 & -0.29 &  --   &  --   &   --  & -0.29$\pm$0.025\\
                       &      &      &       &       & -0.25 & -0.21 & -0.24 & -0.28 &  --   &  --   &   --  & -0.25$\pm$0.020\\
HD\,132475             & 5705 &3.79  & -1.50 &  1.4  & -0.51 & -0.53 & -0.57 &   --  &  --   &  --   &   --  & -0.54$\pm$0.022\\
                       &      &      &       &       & -0.34 & -0.37 & -0.34 &   --  &  --   &  --   &   --  & -0.35$\pm$0.013\\
HD\,140283             & 5725 &3.68  & -2.41 &  1.5  & -1.12 & -1.09 &   --  &   --  &  --   &  --   &   --  & -1.11$\pm$0.015\\
                       &      &      &       &       & -0.64 & -0.63 &   --  &   --  &  --   &  --   &   --  & -0.64$\pm$0.005\\
HD\,160617             & 5940 &3.80  & -1.78 &  1.5  & -1.09 & -1.00 &   --  &   --  &  --   &  --   &   --  & -1.05$\pm$0.045\\
                       &      &      &       &       & -0.72 & -0.68 &   --  &   --  &  --   &  --   &   --  & -0.70$\pm$0.020\\
HD\,166913             & 6050 &4.13  & -1.55 &  1.3  & -0.62 & -0.66 & -0.60 &   --  &  --   &  --   &   --  & -0.63$\pm$0.022\\
                       &      &      &       &       & -0.42 & -0.43 & -0.35 &   --  &  --   &  --   &   --  & -0.40$\pm$0.033\\
HD\,175179             & 5780 &4.18  & -0.74 &  1.0  &  0.03 &  0.02 & -0.01 &  0.00 &  0.03 & -0.01 &   --  &  0.01$\pm$0.017\\
                       &      &      &       &       &  0.10 &  0.09 &  0.12 &  0.03 &  0.10 &  0.06 &   --  &  0.08$\pm$0.026\\
HD\,188510             & 5480 &4.42  & -1.67 &  0.8  & -0.37 & -0.41 & -0.45 &   --  &   --  &   --  &   --  & -0.41$\pm$0.027\\
                       &      &      &       &       & -0.27 & -0.33 & -0.35 &   --  &   --  &   --  &   --  & -0.32$\pm$0.031\\
HD\,189558             & 5670 &3.83  & -1.15 &  1.2  & -0.47 & -0.43 & -0.42 &   --  &   --  &   --  &   --  & -0.44$\pm$0.020\\
                       &      &      &       &       & -0.33 & -0.33 & -0.30 &   --  &   --  &   --  &   --  & -0.32$\pm$0.013\\
HD\,195633             & 6000 &3.86  & -0.64 &  1.4  & -0.12 & -0.18 & -0.13 & -0.13 &   --  &   --  &   --  & -0.14$\pm$0.020\\
                       &      &      &       &       & -0.03 & -0.08 & -0.03 & -0.06 &   --  &   --  &   --  & -0.05$\pm$0.020\\
HD\,205650             & 5815 &4.52  & -1.13 &  1.0  & -0.13 & -0.19 & -0.17 & -0.21 &   --  &   --  &   --  & -0.18$\pm$0.025\\
                       &      &      &       &       & -0.09 & -0.10 & -0.09 & -0.17 &   --  &   --  &   --  & -0.11$\pm$0.029\\
HD\,298986             & 6085 &4.26  & -1.33 &  1.3  & -0.97 & -1.00 &   --  &   --  &   --  &   --  &   --  & -0.99$\pm$0.015\\
                       &      &      &       &       & -0.82 & -0.85 &   --  &   --  &   --  &   --  &   --  & -0.84$\pm$0.015\\
\noalign{\smallskip}\hline
\end{tabular}
\label{table2}
\end{table*}

\subsection{Comparison with other work}

Some groups have determined copper abundances for metal-poor stars based on both LTE and NLTE analysis. We compare our results with those from the previous works, and discuss the reasons for the differences.

\vskip 0.2cm
\noindent{\underline{\cite{NS11}}}
\vskip 0.1cm

\citet{NS11} adopted a LTE line formation for a sample of $\alpha$ rich and poor moderately metal-poor stars, and they used \ion{Cu}{1} $\lambda$ 5105.5, 5218.2 and 5782.1 \AA\ lines determining the copper abundances. Their results have been revised by \citet{YSN16} with the NLTE effects included. Our results are well in agreement with theirs, the average difference of the NLTE results is $0.029\pm0.055$ for the 11 stars in common. Compared with \citet{NS11} our LTE result is $0.023\pm0.040$ lower than theirs, while it is $0.068\pm0.042$ higher for our NLTE results.

\vskip 0.2cm
\noindent{\underline {\citet{MKS02,MGB11}}}
\vskip 0.1cm

Their analysis investigated a large number of metal-poor stars, of which five objects are in common with ours, and an average $\Delta{\rm[Cu/Fe]}$ of $0.132\pm0.139$\,dex is obtained for both LTE results. It is found that the largest difference (0.47\,dex) comes from the metal-poor giant HD\,122563, and around 0.3\,dex difference is due to the continuum scattering included in our analysis. Excluding this object the difference will be reduced to -0.048\,dex.

\vskip 0.2cm \noindent{\underline{\cite{BIR04}}}
\vskip 0.1cm The authors determined the copper abundances for 38 FGK stars. Our LTE results are mostly in agreement with theirs. For the eight stars in common, the average difference is $-0.12\pm0.091$. The largest difference (0.43\,dex) is from the object HD\,166913, both studies have adopted similar stellar parameters, thus it is hard to explain such a large difference in [Cu/Fe] for this star.

\vskip 0.2cm
\noindent{\underline{\citet{SGC91}}}
\vskip 0.1cm

Using high resolution, high signal-to-noise ratio spectra, \citet{SGC91} measured the copper abundances of metal-poor stars. We have three stars in common with this work, one is the giant star HD\,122563, the others are two dwarf stars, i.e., HD\,76932 and HD\,188510. For the giant our LTE copper abundance is 0.26\,dex lower than theirs, most of the difference can be explained by the continuum scattering considered in our analysis. The difference in [Cu/Fe] is 0.04\,dex for HD\,76932, while it is 0.29 for HD\,188510. For the later object the large difference may due to the different stellar parameters adopted in each work.

\vskip 0.2cm \noindent {\underline{\citet{RLA06}}}
\vskip 0.1cm

Based on the high resolution, high signal-to-noise ratio spectra of 176 nearby thick-disk candidate stars, they derived the abundance ratios of [Cu/Fe]. The results of \citet{RLA06} are very much in agreement with ours, the average difference between theirs and our LTE results is $0.06 \pm 0.04$ for the two common stars.

\vskip 0.2cm \noindent {\underline{\citet{ABC17}}}
\vskip 0.1cm

Recently \citet{ABC17} investigated the NLTE effects of copper lines in very metal-poor stars, we have three objects in common with theirs, i.e., HD\,84937, HD\,122563 and HD\,140283. For HD\,84937 their NLTE result is 0.25\,dex higher than ours, while it is 0.38 for HD\,140283. For the giant star HD\,122563 their [Cu/Fe] is 0.69\,dex higher, we note that they have not considered the impact of the continuum scattering, which will result in $\sim$ 0.3\,dex difference, while the rest may due to their large NLTE effects. Similar large NLTE corrections have been found when the hydrogen collisions have not been included (\SH\ = 0.0), and/or the broadening and the exact wavelengths of the two strong \ion{Cu}{1} resonance UV lines at 3247 and 3273 \AA\ have not considered properly.

\vskip 0.2cm \noindent {\underline{\citet{RB18}}}
\vskip 0.1cm

Very recently \citet{RB18} tested the copper abundances in late-type stars using ultraviolet \ion{Cu}{2} lines, and showed that LTE underestimates the copper abundance determined from \ion{Cu}{1} lines, namely the [Cu/H] ratios determined from \ion{Cu}{2} lines are $0.36\pm0.06$\,dex higher than those determined from \ion{Cu}{1} lines for their metal-poor samples. For the four common stars the average difference between their [Cu I/Fe] ratios and ours is $-0.095\pm0.025$, while it is $0.005\pm0.125$ between their [Cu II/Fe] ratios and our NLTE results. The difference of [Cu I/Fe] for the two works is due to the lower $log~gf$ values for the two resonance lines ( about 0.15\,dex lower) adopted by us.

\section {Discussion}

Following \citet{NS10} we classified our samples as halo (h), thick disk or halo (tk/h), thick disk (tk), and thin disk (tn) stars, and presented in Table \ref{table3}. In Fig. \ref{fig5} we plotted the Toomre diagram of our program stars. The behavior of [Cu/Fe] with the stellar metallicity [Fe/H] holds information about the chemical evolution of our Galaxy. Figs. \ref{fig6} and \ref{fig7} display the trend of the [Cu/Fe] ratio (calculated in LTE and NLTE) with the metal abundance for all stars investigated in this paper, respectively. One important feature that we can find from Fig. \ref{fig7} is that the [Cu/Fe] ratios decrease with decreasing metallicity for [Fe/H] from $\sim-1.0$ to $\sim-2.5$, and they may increase for more metal-poor stars. While \citet{ABC17} suggested that the trend of [Cu/Fe] ratio is constant, this is due to their very large NLTE effects ($\sim$1\,dex) for very metal-poor stars. Another important feature that can be seen from this figure is that there is a group of stars with clear low [Cu/Fe] ratios compared with similar metallicity other samples, and it is interesting to discuss the behavior of this group of stars. It is first noted by \citet{NS11} that the low-$\alpha$ members show systematic low copper abundances \citep[also see][]{YSN16}, and in our sample there are four such type objects from their work, i.e. CD\,$-57^\circ$ 1633, HD\,103723, HD\,122196 and HD\,298986, the [Cu/Fe] ratios of these four stars are obviously lower.

\begin{figure}
\resizebox{\hsize}{6.6cm}{\includegraphics[width=9.3cm]{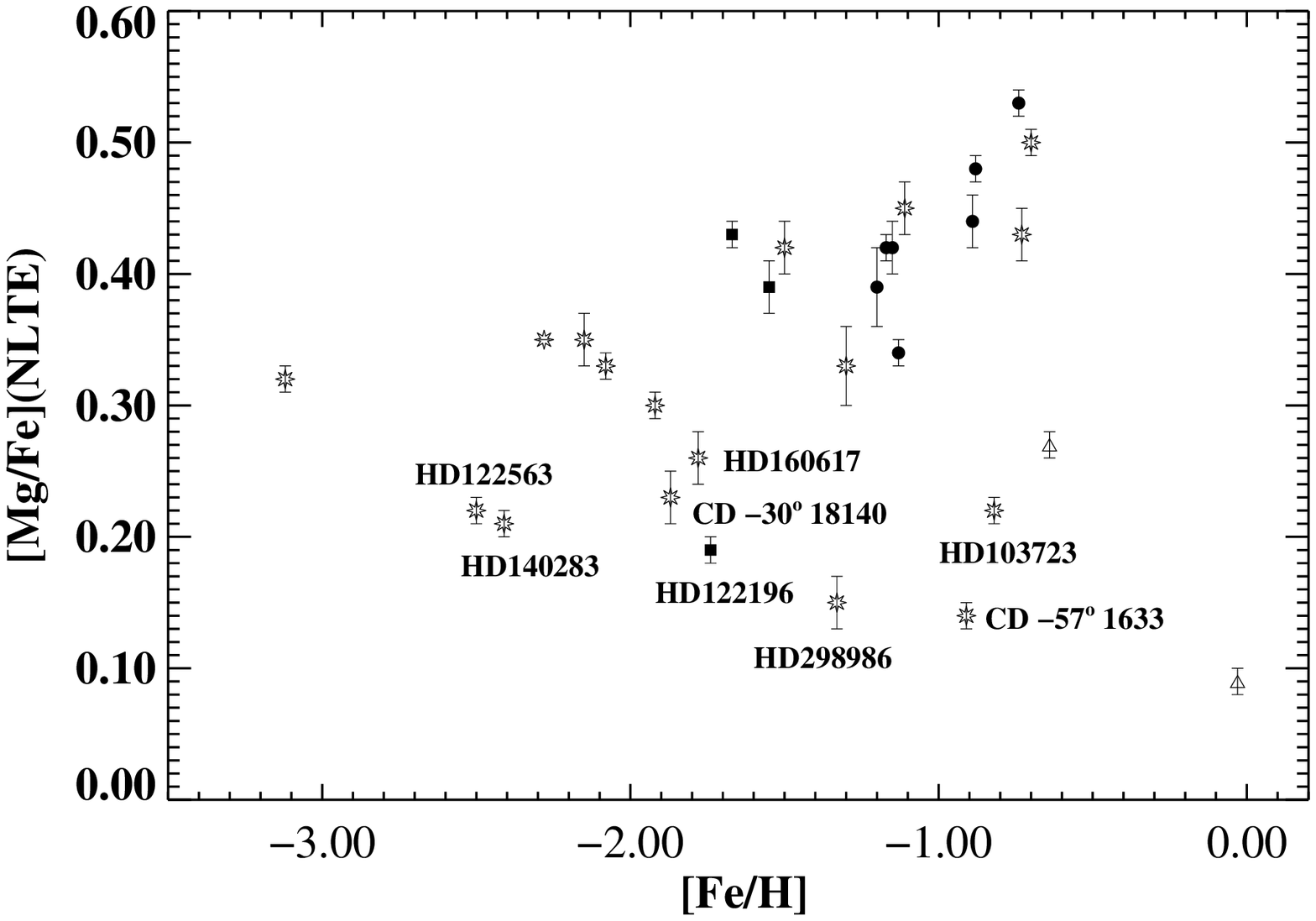}}
\resizebox{\hsize}{6.6cm}{\includegraphics[width=9.3cm]{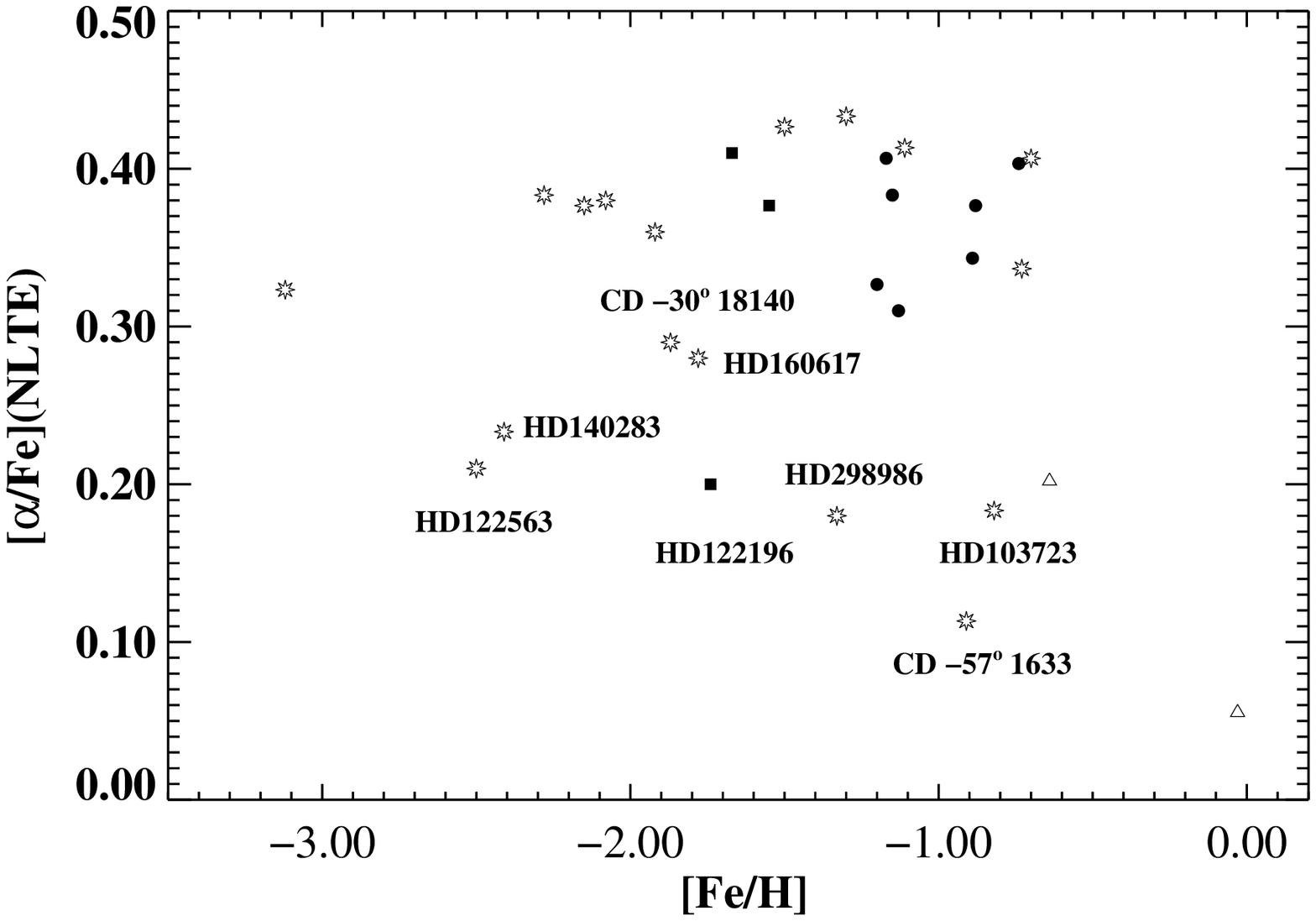}}
\caption[short title]{The [Cu/Fe] ratios under NLTE situation as a function of [Fe/H] for selected stars. Symbols are same as Fig. \ref{fig5}.} \label{fig8}
\end{figure}

In order to investigate the behavior of $\alpha$-elments for our sample stars, we derived the LTE and NLTE Mg, Si and Ca abundances of those objects based on the atomic models of magnesium \citep{ZSP17}, silicon \citep{ZSP16} and calcium \citep{Ma17}, and the detailed information on abundance and kinematics was listed in Table \ref{table3}. This table shows that, besides above four low copper abundance stars, the ratios of both [Mg/Fe] and [$\alpha$/Fe] (the average of Mg, Si and Ca abundances) for the other four such type halo stars (i.e. CD\,$-30^\circ$ 18140, HD\,122563, HD\,140283 and HD\,160617) are also obviously lower compared to other normal stars with similar metallicity, which can also be clear seen in the plots of [Mg/Fe] and [$\alpha$/Fe] versus metallicity in Fig. \ref{fig8}. For the two very metal-stars, HD\,140283 and HD\,122563, \citet{SAB15} revealed that both are excellent examples of abundances dominated by the weak r-process. 

\begin{figure}
\resizebox{\hsize}{6.6cm}{\includegraphics[width=9.3cm]{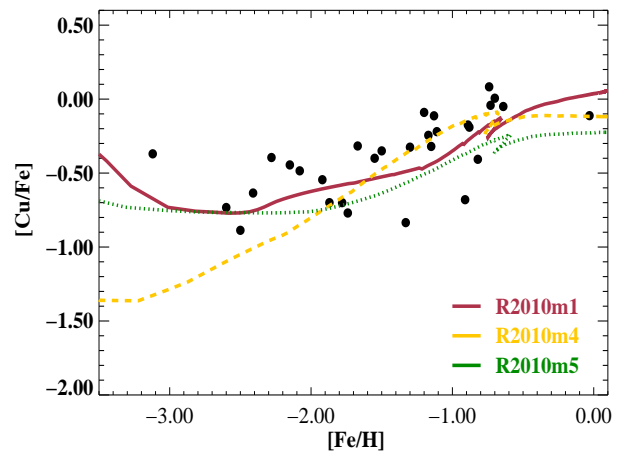}}
\caption[short title]{Comparison with the Galactic chemical evolution models from \citet{RKT10}, and it seems that their model 1 can reproduce our results.} \label{fig9}
\end{figure}

Detailed modeling of the Galactic chemical evolution for copper has been attempted by many authors \citep[e.g.][]{SGC91,MRB93,TWW95,GP00,KUN06,RM07,RKT10}. \citet{SGC91} suggested that the copper contributes mainly from the weak s-process, however, \citet{RGB92} argued that a large fraction is from long lived type Ia supernovae \citep[also see][]{MRB93,MKS02}. Based on \cite{WW95} metallicity-dependent yields, \citet{TWW95} calculated the behavior of [Cu/Fe] as a function of metallicity, and their result predicted that this elements may be synthesized in significant amounts by the nuclear burning stages in massive stars, which was confirmed by \citet{RM07} with a comprehensive study of copper evolution in different systems. In Fig. \ref{fig9} we plot the predicted [Cu/Fe] rations as a function of [Fe/H] from \citet{RKT10} with our NLTE results. In their \emph{ model 1}, the \cite{WW95} case B yields for normal SNe II have been adopted and it provides the best fit to the observed [Cu/Fe] versus [Fe/H] trend, even for the lowest metallicities, which means that copper is mainly produced by massive type II SNe. The models 4 and 5 of \citet{RKT10} were computed adopting the \citet{KUN06} yields with $\varepsilon_{\rm{HN}}$ = 0 and 1, respectively. Although model 4 can reproduce the [Cu/Fe] ratio for metal-rich stars, it underestimate the copper abundance for the very metal-poor region. Both models 1 and 5 are indistinguishable with the observational data alone, though the predicted [Cu/Fe] ratio of model 5 is slightly lower than that of the observed one.

\section{Conclusions}

We have determined copper abundances for 29 meal-poor stars spanning the metallicity range $-3.2<$[Fe/H]$<\sim$0.0\,dex. Using the MAFAGS's LTE model atmospheres the copper abundances were obtained with both the near-UV and optical lines. Our results are derived for both LTE and NLTE based on the line fitting method. It is found:
\begin{table*}
\caption[3]{The LTE and NLTE magnesium, silicon and calcium abundances of our sample stars.}
\setlength{\tabcolsep}{0.1cm}
\centering
\begin{tabular}{lccccccrrrc}
\hline\hline\noalign{\smallskip}
   Name   & [Mg/Fe]   &  [Mg/Fe] & [Si/Fe]  & [Si/Fe] & [Ca/Fe] & [Ca/Fe] &  U   &  V   &   W  & pop\footnote{Following \citet{NS10} classification as halo (h), thick disk or halo (tk/h), thick disk (tk), and thin disk (tn) stars.}\\
\hline\noalign{\smallskip}
          &  LTE      &  NLTE    &  LTE     &  NLTE   &  LTE    &  NLTE   &$km/s$&$km/s$&$km/s$& \\
\noalign{\smallskip}\hline\noalign{\smallskip}
CD\,$-$30$^\circ$ 18140&  0.14$\pm$0.01 & 0.23$\pm$0.02 & 0.25$\pm$0.00 & 0.26$\pm$0.00 &  0.27$\pm$0.01  & 0.38$\pm$0.02&  71.5&-195.5& -11.2&h\\
CD\,$-$57$^\circ$ 1633 &  0.06$\pm$0.00 & 0.14$\pm$0.01 & 0.10$\pm$0.04 & 0.09$\pm$0.01 &  0.08$\pm$0.02  & 0.11$\pm$0.02&-329.0&-240.0& -28.0&h\\
G\,13-009              &  0.26$\pm$0.05 & 0.35$\pm$0.00 & 0.19$\pm$0.15 & 0.33$\pm$0.03 &  0.30$\pm$0.03  & 0.47$\pm$0.03& -93.1&-264.4& -81.4&h\\
G\,020-024             &  0.22$\pm$0.01 & 0.30$\pm$0.01 & 0.26$\pm$0.05 & 0.34$\pm$0.03 &  0.31$\pm$0.02  & 0.44$\pm$0.03& 160.3&-206.0& 64.3&h\\
G\,64-12               &  0.22$\pm$0.01 & 0.32$\pm$0.01 & 0.04$\pm$0.00 & 0.28$\pm$0.00 &  0.31$\pm$0.00  & 0.37$\pm$0.00& -50.0&-317.0&397.0&h\\
G\,183-011             &  0.26$\pm$0.04 & 0.33$\pm$0.01 & 0.36$\pm$0.00 & 0.41$\pm$0.00 &  0.27$\pm$0.02  & 0.40$\pm$0.03&  53.5&-379.0&-31.6&h\\
HD\,61421              &  0.04$\pm$0.02 & 0.09$\pm$0.01 & 0.15$\pm$0.09 & 0.08$\pm$0.01 &  0.01$\pm$0.01  & 0.00$\pm$0.00& -12.6&   7.0&  6.4&tn\\
HD\,76932              &  0.33$\pm$0.02 & 0.44$\pm$0.02 & 0.36$\pm$0.06 & 0.33$\pm$0.02 &  0.25$\pm$0.02  & 0.26$\pm$0.02&   0.3& -41.0&-77.0&tk\\
HD\,84937              &  0.28$\pm$0.04 & 0.35$\pm$0.02 & 0.24$\pm$0.07 & 0.37$\pm$0.00 &  0.35$\pm$0.02  & 0.41$\pm$0.02& 226.0&-237.5& -8.4&h\\
HD\,97320              &  0.28$\pm$0.02 & 0.39$\pm$0.03 & 0.38$\pm$0.04 & 0.36$\pm$0.01 &  0.17$\pm$0.02  & 0.23$\pm$0.02&  80.0& -11.0& -30.0&tk\\
HD\,97916              &  0.39$\pm$0.02 & 0.48$\pm$0.01 & 0.42$\pm$0.05 & 0.40$\pm$0.02 &  0.23$\pm$0.03  & 0.25$\pm$0.02& 117.7&  15.9& 96.1&tk\\
HD\,103723             &  0.13$\pm$0.01 & 0.22$\pm$0.01 & 0.19$\pm$0.05 & 0.17$\pm$0.01 &  0.14$\pm$0.01  & 0.16$\pm$0.03& -72.0&-193.0& 58.0&h\\
HD\,106038             &  0.24$\pm$0.03 & 0.33$\pm$0.03 & 0.76$\pm$0.05 & 0.75$\pm$0.01 &  0.16$\pm$0.02  & 0.22$\pm$0.02&  25.2&-264.3& 26.2&h\\
HD\,111980             &  0.35$\pm$0.00 & 0.45$\pm$0.02 & 0.47$\pm$0.03 & 0.45$\pm$0.01 &  0.31$\pm$0.01  & 0.34$\pm$0.03& 239.0&-174.0& 57.0&h\\
HD\,113679             &  0.42$\pm$0.00 & 0.50$\pm$0.01 & 0.41$\pm$0.03 & 0.39$\pm$0.01 &  0.35$\pm$0.02  & 0.33$\pm$0.03& -96.0&-278.0&  3.0&h\\
HD\,121004             &  0.36$\pm$0.02 & 0.43$\pm$0.02 & 0.40$\pm$0.03 & 0.40$\pm$0.01 &  0.17$\pm$0.10  & 0.18$\pm$0.08&  70.0&-242.0&105.0&h\\
HD\,122196             &  0.11$\pm$0.01 & 0.19$\pm$0.01 & 0.18$\pm$0.01 & 0.17$\pm$0.02 &  0.16$\pm$0.03  & 0.24$\pm$0.04&-160.9&-139.4& 23.5&tk/h\\
HD\,122563             &  0.17$\pm$0.01 & 0.22$\pm$0.01 & 0.25$\pm$0.02 & 0.24$\pm$0.00 &  0.07$\pm$0.02  & 0.17$\pm$0.01& 139.0&-233.0& 26.0&h\\
HD\,126681             &  0.36$\pm$0.01 & 0.42$\pm$0.01 & 0.46$\pm$0.02 & 0.45$\pm$0.01 &  0.34$\pm$0.03  & 0.35$\pm$0.01& -15.0& -28.0&-64.0&tk\\
HD\,132475             &  0.34$\pm$0.02 & 0.42$\pm$0.02 & 0.55$\pm$0.04 & 0.54$\pm$0.01 &  0.25$\pm$0.02  & 0.32$\pm$0.04& 44.0&-371.0&60.0&h\\
HD\,140283             &  0.13$\pm$0.03 & 0.21$\pm$0.01 & 0.16$\pm$0.02 & 0.17$\pm$0.02 &  0.08$\pm$0.04  & 0.25$\pm$0.04& 30.0& 147.0&-320.5&h\\
HD\,160617             &  0.18$\pm$0.01 & 0.26$\pm$0.02 & 0.26$\pm$0.04 & 0.26$\pm$0.02 &  0.21$\pm$0.02  & 0.32$\pm$0.03& 68.2&-209.5&-85.7&h\\
HD\,166913             &  0.30$\pm$0.00 & 0.39$\pm$0.02 & 0.40$\pm$0.07 & 0.39$\pm$0.04 &  0.27$\pm$0.02  & 0.35$\pm$0.02& -44.3& -44.5& 67.6&tk/h\\
HD\,175179             &  0.45$\pm$0.01 & 0.53$\pm$0.01 & 0.41$\pm$0.04 & 0.40$\pm$0.01 &  0.36$\pm$0.03  & 0.34$\pm$0.02& 96.0&-102.0&-19.0&tk\\
HD\,188510             &  0.36$\pm$0.01 & 0.43$\pm$0.01 & 0.37$\pm$0.00 & 0.38$\pm$0.00 &  0.36$\pm$0.03  & 0.42$\pm$0.02&-141.1&-109.4& 71.6&tk/h\\
HD\,189558             &  0.34$\pm$0.01 & 0.42$\pm$0.02 & 0.41$\pm$0.04 & 0.40$\pm$0.01 &  0.30$\pm$0.03  & 0.33$\pm$0.03& 89.0&-109.0&48.0&tk\\
HD\,195633             &  0.19$\pm$0.02 & 0.27$\pm$0.01 & 0.25$\pm$0.06 & 0.22$\pm$0.02 &  0.11$\pm$0.03  & 0.12$\pm$0.03& -48.3& -15.7& -3.5&tn\\
HD\,205650             &  0.26$\pm$0.01 & 0.34$\pm$0.01 & 0.36$\pm$0.03 & 0.36$\pm$0.02 &  0.20$\pm$0.01  & 0.23$\pm$0.02&-114.0&-71.0 & 18.0&tk\\
HD\,298986             &  0.06$\pm$0.03 & 0.15$\pm$0.02 & 0.21$\pm$0.06 & 0.20$\pm$0.04 &  0.12$\pm$0.01  & 0.19$\pm$0.02& 250.1&-138.0&163.3&h\\
\noalign{\smallskip}\hline
\noalign{\smallskip}\hline
\end{tabular}
\label{table3}
\end{table*}

\begin{enumerate}
\item The [Cu/Fe] ratios are under-abundant for metal-poor stars, and there is an indication that [Cu/Fe] decreases with decreasing metallicity within $\sim-2.0<$ [Fe/H] $<\sim -0.7$. While, it may increase for very metal-poor stars with [Fe/H] $<$ -3.0, which need to be confirmed with more objects.
\item Our NLTE result confirms that the low $\alpha$ sample stars are also with lower copper abundance found by \citet{NS11}.
\item The NLTE effects of \ion{Cu}{1} lines are sensitive to the metallicity, and they increase with decreasing metallicity. The NLTE effects can reach $\sim$ 0.5\,dex for very metal-poor stars, which can explain the large difference of copper abundance derevied from \ion{Cu}{2} and \ion{Cu}{1} resonance lines noted by \citep{RSL14,RB18}. It is also need to be pointed out that the NLTE effects are different from line to line, and the weak lines are less sensitive to NLTE effects, while the strong resonance 3247 and 3273 \AA\ lines show large NLTE effects.
\item  Compared with the \emph{model 1} of \citet{RKT10} our results suggest that, similar as $\alpha$ elements, copper is mainly produced by massive type II SNe.
\end{enumerate}

Our results indicate that it is important to perform NLTE abundance analysis for \ion{Cu}{1} lines for very metal-poor stars.

\begin{acknowledgements}
J.R. acknowledges Dr. Aoki for providing the Subaru spectrum for G64-12. This research was supported by the National Key Basic Research Program of China under grant No. 2014CB845700 and the National Natural Science Foundation of China under grant Nos. 11473033 and 11603037. This work is also supported by the Astronomical Big Data Joint Research Center, co-founded by the National Astronomical Observatories, Chinese Academy of Sciences and the Alibaba Cloud.

\end{acknowledgements}


\begin{thebibliography}{}
\bibitem[Andrievsky et al.(2017)]{ABC17} Andrievsky, S., Bonifacio, P., Caffau, E., Korotin,S., et al. 2018, \mnras, 473, 3377
\bibitem[Anstee \& O'Mara(1991)]{AO91}
      Anstee, S. D., O'Mara, B. J., 1991, \mnras, 253, 549
\bibitem[Anstee \& O'Mara(1995)]{AO95}
      Anstee, S. D.,\& O'Mara, B. J., 1995, \mnras, 276, 859
\bibitem[Arlandini et~al.(1999)]{AKW99}
      Arlandini, C., K$\ddot{a}$ppeler, F., Wisshak, K., et al. 1999, \apj, 525, 886
\bibitem[Asplund et~al.(2009)]{AGS09}
      Asplund, M., Grevesse, N., Sauval, A. J., \& Scott, P. 2009, \araa, 47, 481
\bibitem[Bagbulo et~al.(2005)]{BJL05}
      Bagbulo, S., Jehin E., \& Ledoux C., et al. 2005, ESO messenger, 114, 10
\bibitem[Biehl (1976)]{B76}
      Biehl, D., PhD thesis, Univ. Kiel, 1976
\bibitem[Bihain et~al.(2004)]{BIR04}
      Bihain, G., Israelian, G., Rebolo, R., Bonifacio, P., \& Molaro, P., 2004, \aap, 423, 777
\bibitem[Bonifacio, Caffau \& Ludwig(2010)]{BCH10}
      Bonifacio, P., Caffau, E., \& Ludwig, H.-G. 2010, \aap, 524, 96
\bibitem[Butler \& Giddings(1985)]{BG85}
      Butler, K., \& Giddings J. 1985, Newsletter on the analysis os astronomical
      spectra No. 9, University of London
\bibitem[Canuto \& Mazzitelli(1992)]{CM92}
      Canuto, V. M., \& Mazzitelli, I. 1992, \apj, 389, 724
\bibitem[Carretta et~al.(2010)]{CBG10}
      Carretta, E., Bragaglia, A., Gratton, R. G., et al. 2010, \aap, 520, A95
\bibitem[Carretta et~al.(2014)]{CBG14}
      Carretta, E., Bragaglia, A., Gratton, R. G., et al., 2014, \aap, 564, A60
\bibitem[Carretta et~al.(2015)]{CBG15}
      Carretta, E., Bragaglia, A., Gratton, R. G., et al., 2015, \aap, 578, A166
\bibitem[Cayrel et~al.(2004)]{CDS04}
      Cayrel, R., Depagne, E., Spite, M., Hill, V., Spite, F. et al. 2004, \aap, 416, 1117
\bibitem[Colucci et~al.(2012)]{CBC12}
        Colucci1, J. E., Bernstein, R. A., Cameron, S. A., \& McWilliam, A. 2012, \apj, 746, 29
\bibitem[Cunha et~al.(2002)]{CSS02}
      Cunha, K., Smith, V. V., Suntzeff, N. B., et al. 2002, \aj, 124, 379
\bibitem[Delgado Mena et~al.(2017)]{DTA17}
      Delgado Mena, E., Tsantaki, M., Adibekyan, V. Zh., Sousa, S. G., Santos, N. C., Gonz\'{a}lez Hern\'{a}ndez, J. I., \& Israelian, G., 2017,  \aap, 606, A94
\bibitem[Fink et~al.(2014)]{FKS14}
      Fink, M., Kromer, M., Seitenzahl, I. R., et al., 2014, \mnras, 438, 1762
\bibitem[Gaia Collaboration et~al.(2016)]{Gaia16}
      Gaia Collaboration, Prusti, T., de Bruijne, J. H. J., et al. 2016, \aap, 595, A1
\bibitem[Gehren et~al.(2004)]{GLS04}
      Gehren, T., Liang, Y. C., Shi, J. R., et al. 2004, \aap, 413, 1045
\bibitem[Goswami \& Prantzos(2000)]{GP00}
      Goswami A., \& Prantzos N., 2000, \aap, 359, 191
\bibitem[Gratton \& Sneden(1988)]{GS88}
      Gratton, R. G., \& Sneden, C., A\&A, 204,193
\bibitem[Grupp(2004)]{G04}
      Grupp, F. 2004, \aap, 420, 289
\bibitem[Grupp, Kurucz \& Tan(2009)]{GKT09}
      Grupp, F., Kurucz, R. L., \& Tan, K. F. 2009, \aap, 530, 177
\bibitem[Ishigaki, Aoki \& Chiba(2013)]{IAC03}
      Ishigaki, M. N., Aoki, W., \& Chiba, M., 2013, \apj, 771, 67
\bibitem[Iwamoto et~al.(1999)]{IBN99}
      Iwamoto, K., Brachwitz, F., Nomoto, K. et al. 1999, \apjs, 125, 439
\bibitem[Johnson, Ivans, \& Stetson (2006)]{JIS06}
      Johnson, J. A., Ivans, I. I., \& Stetson, P. B., 2006, \apj, 640, 801
\bibitem[Johnson et~al.(2014)]{JRK14}
      Johnson, C. I., Rich, R. M., Kobayashi, C., Kunder, A., \& Koch, A., 2014, \aj, 148, 67
\bibitem[Kobayashi et~al.(2006)]{KUN06}
      Kobayashi, C., Umeda H., Nomoto, K., Tominaga, N., Ohkubo, T., 2006, \apj, 653, 1145
\bibitem[Koch \& McWilliam(2014)]{KM14}
      Koch, A., \& McWilliam, A., 2014, \aap, 565, A23
\bibitem[Mucciarelli et al.(1993)]{MRB93}
      Matteucci, F., Raiteri, C. M., Busson, M., Gallino, R., \& Gratton, R., 1993, \aap, 272, 421
\bibitem[Lai et~al.(2008)]{LBJ08}
      Lai, D. K., Bolte, M., Johnson, J. A., Lucatello, S., Heger, A., \& Woosley, S.E., 2008, \apj, 681, 1524
\bibitem[Limongi \& Chieffi (2003)]{LC03}
      Limongi, M., \& Chieffi, A., 2003, \apj, 592, 404
\bibitem[Liu et~al.(2014)]{LCZ14}
      Liu, Y. P., Gao, C., Zeng, J. L., Yuan, J. M., \& Shi, J. R. 2014, \apjs, 211, 30
\bibitem[Mashonkina et al.(2011)]{MGS11}
      Mashonkina, L., Gehren, T., Shi, J. R., Korn, A. J., \& Grupp, F., 2011, \aap, 528, A87
\bibitem[Mashonkina et al.(2017)]{MJP17}
      Mashonkina, L., Jablonka, P., Pakhomov, Yu., Sitnova, T., \& North, P., 2017, \aap, 604, A129
\bibitem[Mashonkina et al.(2017)]{Ma17}
      Mashonkina, L., Sitnova, T., \& Belyaev, A. K., 2017, \aap, 605, A53
\bibitem[McWilliam (2016)]{M16}
      McWilliam, A., 2016, \pasp, 33, 40
\bibitem[McWilliam \& Smecker-Hane(2005)]{MS05}
      McWilliam, A. \& Smecker-Hane, T., A., 2005, \apjl, 622, 29
\bibitem[McWilliam et~al.(2013)]{MWM13}
      McWilliam, A., Wallerstein, G., \& Mottini, M., 2013, \apj, 778, 149
\bibitem[Mikolaitis et al.(2017)]{MLR17}
      Mikolaitis, \v{S}., de Laverny, P., Recio¨CBlanco, A., Hill, V., Worley, C. C., \& de Pascale, M., 2017, \aap, 600, A22
\bibitem[Mishenina et~al.(2011)]{MGB11}
      Mishenina, T. V., Gorbaneva, T. I., Basak, N. Yu., Soubiran, C., \& Kovtyukh, V. V., 2011, Astronomy Reports, 55, 680
\bibitem[Mishenina et~al.(2002)]{MKS02}
      Mishenina, T. V., Kovtyukh, V. V., Soubiran, C., Travaglio, C., \&  Busso, M., 2002, \aap, 396, 189
\bibitem[Nissen \& Schuster(2010)]{NS10}
      Nissen, P. E., \& Schuster, W. J., 2010, \aap, 511, L10
\bibitem[Nissen \& Schuster(2011)]{NS11}
      Nissen, P. E., \& Schuster, W. J., 2011, \aap, 530, A15
\bibitem[Noguchi, Aoki \& Kawanomoto(2002)]{NAK02}
      Noguchi, K., Aoki, W., \& Kawanomoto, S., 2002,  \pasj, 54, 855
\bibitem[Pancino et al.(2002)]{PPH02}
      Pancino, E., Pasquini, L., Hill, V., Ferraro, F. R., \& Bellazzini, M., \apjs, 568, 101
\bibitem[Pignatari et al.(2010)]{PGH10}
      Pignatari, M., Gallino, R., Heil, M., Wiescher, M., K{\"a}ppeler, F., Herwig, F. and Bisterzo, S., \apj, 710, 1557
\bibitem[Pomp\'{e}ia et al.(2008)]{PHS08}
      Pomp\'{e}ia, L., Hill, V., Spite, M., et al. 2008, \aap, 480, 379
\bibitem[Primas et~al.(2000)]{PBS00}
      Primas, F., Brugamyer, E., Sneden, C., et al. 2000, LIACo, 35, 119
\bibitem[Reddy \& Lambert(2008)]{RL08}
      Reddy, B. E., \& Lambert, D. L., 2008, \mnras, 391, 95
\bibitem[Reddy et~al.(2006)]{RLA06}
      Reddy, B. E., Lambert, D. L., Allende Prieto, C., \mnras, 367, 1329
\bibitem[Reddy et~al.(2003)]{RTL03}
      Reddy, B. E., Tomkin, J., Lambert, D. L., \& Allende Prieto, C., 2003, \mnras, 340, 304
\bibitem[Reetz (1991)]{R91}
      Reetz, J. K. 1991, PhD thesis, Universit\"{a}t M\"{u}nchen
\bibitem[Raiteri, Gallino \& Busso (1992)]{RGB92}
      Raiteri, C. M., Gallino, R., \& Busso, M., 1992, \apj, 387, 263
\bibitem[Roederer \& Barklem(2018)]{RB18}
      Roederer, I. U., \& Barklem, P.S., 2018, \apj, 857, 2
\bibitem[Roederer \& Lawler(2012)]{RL12}
      Roederer, I. U., \& Lawler, J.E., 2012, \apj, 750, 76
\bibitem[Roederer et al.(2014)]{RSL14}
      Roederer, I. U., Schatz, H., Lawler, J. E., et al. 2014, \apj, 791, 32
\bibitem[Romano et al. (2010)]{RKT10}
      Romano, D., Karakas, A. I., Tosi, M., \& Matteucci, F., \aap, 2010, 522, A32
\bibitem[Romano \& Matteucci (2007)]{RM07}
      Romano, D., \& Matteucci, F., 2007, \mnras, 378, L59
\bibitem[Rybicki \& Hummer(1991)]{RH91}
      Rybicki, G. B., \& Hummer, D. G. 1991, \aap, 245, 171
\bibitem[Rybicki \& Hummer(1992)]{RH92}
      Rybicki, G. B., \& Hummer, D. G., 1992, \aap, 262, 209
\bibitem[Sakari, McWilliam \& Wallerstein(2017)]{SMW17}
      Sakari, C. M., McWilliam, A., \& Wallerstein, G., 2017, arXiv:1701.03802
\bibitem[Shetrone et~al.(2003)]{SVT03}
      Shetrone, M., Venn, K. A., Tolstoy, E., Primas, F., Hill, V., \& Kaufer, A., 2003, \aj, 125, 684
\bibitem[Shi et~al.(2009)]{SGM09}
      Shi, J. R., Gehren, T., Mashonkina, L., \& Zhao, G., 2009, A\&A, 503, 533
\bibitem[Shi et~al.(2014)]{SGZ14}
      Shi, J. R., Gehren, T., Zeng, J. L., Mashonkina, L., \& Zhao, G. 2014, ApJ, 782, 80
\bibitem[Shi, Gehren \& Zhao(2004)]{SGZ04}
      Shi, J. R., Gehren, T., \& Zhao, G., 2004, \aap, 423, 683
\bibitem[Simmerer et~al.(2003)]{SSI03}
      Simmerer, J., Sneden, C., Ivans, I. I., Kraft, R. P., Shetrone, M. D., \& Smith, V. V., 2003, \aj, 125,2018
\bibitem[Siqueira-Mello et~al.(2015)]{SAB15}
      Siqueira-Mello, C., Andrievsky, S. M., Barbuy, B., Spite, M., Spite, F., \& Korotin, S. A., 2015, \aap, 584, A86
\bibitem[Smith et~al.(2000)]{SSC00}
      Smith, V. V., Suntzeff, N. B., Cunha, K., et al., 2000, \aj, 119, 1239
\bibitem[Sneden \& Crocker(1988)]{SC88}
      Sneden, C., \& Crocker, D. A., \apj, 335, 406
\bibitem[Sneden et al.(1991)]{SGC91}
      Sneden, C., Gratton, R. G., \& Crocker, D. A., \aap, 246, 354
\bibitem[Sbordone et~al.(2007)]{SBB07}
      Sbordone, L., Bonifacio, P., Buonanno, R., et al. 2007, \aap, 465, 815
\bibitem[Tan, Shi \& Zhao(2009)]{TSZ09}
      Tan, K. F., Shi, J. R., \& Zhao, G., 2009, \mnras, 392, 205
\bibitem[Timmes, Woosley \& Weaver(1995)]{TWW95}
      Timmes, F. X., Woosley, S. E., \& Weaver, T. A., 1995, \apjs, 98, 617
\bibitem[Travaglio et al.(2004)]{THR04}
      Travaglio, C., Hillebrandt, W., Reinecke, M., Thielemann, F. -K., \aap, 2004, 425, 1029
\bibitem[Villanova et~al.(2013)]{VGC13}
      Villanova, S., Geisler, D., Carraro, G.; Moni Bidin, C., \& Mu\~{n}oz, C., 2013, \apj, 778, 186
\bibitem[Villanova et~al.(2017)]{VMM17}
      Villanova, S., Moni Bidin, C., Mauro, F., Munoz, C., \& Monaco, L., 2017, \mnras, 464, 2730
\bibitem[Woosley \& Weaver(1995)]{WW95}
      Woosley, S. E., \& Weaver, T. A., 1995, \apjs, 101, 181
\bibitem[Yan et al. (2016)]{YSN16}
      Yan, H. L., Shi, J. R., Nissen, P. E., \& Zhao, G., 2016, \aap, 585, A102
\bibitem[Yan, Shi \& Zhao(2015)]{YSZ15}
      Yan, H. L., Shi, J. R., \& Zhao, G., 2015, \apj, 802, 36
\bibitem[Zhao et al. (2016)]{ZG16}
      Zhao, G., Mashonkina, L., Yan, H. L., Alexeeva, S., Kobayashi, C., et al., 2016, \apj, 833, 225
\bibitem[Zhang et al.(2016)]{ZSP16}
      Zhang, J. B., Shi, J. R., Pan, K. K., Allende Prieto, C., \& Liu, C., 2016, \apj, 833, 137
\bibitem[Zhang et al.(2017)]{ZSP17}
      Zhang, J. B., Shi, J. R., Pan, K. K., Allende Prieto, C., \& Liu, C., 2017, \apj, 835, 90
\end{thebibliography}
\end{document}